\documentclass[aps,prx,twocolumn,preprintnumbers,superscriptaddress]{revtex4}
\usepackage{amsmath}
\usepackage{amssymb}
\usepackage{enumitem}
\usepackage{color}
\usepackage{xcolor}
\usepackage{graphicx}
\usepackage{dcolumn}
\usepackage{bm}
\usepackage[colorlinks=true,linkcolor=blue,citecolor=blue,urlcolor=blue]{hyperref}

\newcommand{\eps}{\epsilon}

\newcommand{\drop}[1]{}

\begin{document}
\title{On Green's function embedding using sum-over-pole representations}

%
%
\author{Andrea \surname{Ferretti}}
\email[corresponding author: ]{andrea.ferretti@nano.cnr.it}
\affiliation{Centro S3, CNR--Istituto Nanoscienze, 41125 Modena, Italy}
\author{Tommaso \surname{Chiarotti}}
\affiliation{Theory and Simulations of Materials (THEOS)
             and National Centre for Computational Design and Discovery of Novel Materials (MARVEL),
           \'Ecole Polytechnique F\'ed\'erale de Lausanne, 1015 Lausanne, Switzerland}
\author{Nicola \surname{Marzari}}
\affiliation{Theory and Simulations of Materials (THEOS)
             and National Centre for Computational Design and Discovery of Novel Materials (MARVEL),
           \'Ecole Polytechnique F\'ed\'erale de Lausanne, 1015 Lausanne, Switzerland}

%
%
\date{\today}

%
%
\begin{abstract}
In Green's function theory, the total energy of an interacting many-electron system can be expressed in a variational form using the Klein or Luttinger-Ward functionals. 
Green's function theory also naturally addresses the case where the interacting system is embedded into a bath. 
This latter can then act as a dynamical (i.e., frequency-dependent) potential, providing a more general framework than that of conventional static external potentials. 
Notably, the Klein functional includes a term of the form $\text{Tr}_\omega \text{Ln}\left\{G_0^{-1}G\right\}$, where $\text{Tr}_\omega$ is the frequency integration of the trace operator. Here, we show that using a sum-over-pole representation for the Green's functions and the algorithmic-inversion method one can obtain in full generality an explicit analytical expression for $\text{Tr}_\omega \text{Ln}\left\{G_0^{-1}G\right\}$. This allows one, e.g., to derive a variational expression for the Klein functional 
in the presence of an embedding bath, or to provide an explicit expression of the RPA correlation energy in the framework of the optimized effective potential.

\end{abstract}

%
%
\maketitle

\section{Introduction}
\label{sec:intro}

Electronic-structure simulations based on density-functional theory (DFT)~\cite{Kohn1999RevModPhys} 
are today widely exploited~\cite{Pribram-Jones2015AnnuRevPhysChem}
in condensed-matter physics, quantum chemistry, or materials modelling~\cite{Marzari2021NatMater}. 
Even if DFT can in principle be used to access any observable of an interacting system as a functional of the density~\cite{Hohenberg-Kohn1964PhysRev,Martin-Reining-Ceperley2016book,Marzari2021NatMater}, currently available functionals and approximations are mostly limited to the ground-state total energy (and, in turn, to its derivatives wrt external parameters) and to observables connected to the charge density. Instead, electronic excitations are typically addressed by extensions of the basic theory, such as time-dependent DFT~\cite{Runge-Gross1984PhysRevLett,Petersilka1996PhysRevLett,Ullrich2012book} or ensemble DFT~\cite{Gross-Oliveira-Kohn1988PRA,Gould2019PhysRevLett,Cernatic2021TopCurrChem}.
Notably, all these approaches are equipped with a variational principle which allows one to determine the basic quantity of the theory (e.g. the density in DFT or its time-dependent version in TD-DFT) for the systems studied. 

Conversely, Green's function (GF) methods~\cite{Stefanucci-vanLeeuwen2013book, Martin-Reining-Ceperley2016book} such as the GW approximation and its combination with the Bethe-Salpeter equation (BSE)~\cite{Hedin1965PR, Reining2018wcms, Golze2019FIC, Onida2002RMP}, are commonly used to address charged and neutral excitations. 
Nevertheless,
the one-particle GF can also be used to access the ground-state total energy~\cite{fett-wale71book, Martin-Reining-Ceperley2016book} (e.g., via the Galitskii-Migdal expression). 
Variationality of the total energy wrt the one-particle Green's function can be recovered by using the Luttinger-Ward or Klein (LWK) functionals~\cite{Luttinger-Ward1960PR,Klein1961PR,Baym1961PhysRev,Almbladh1999InternationalJournalofModernPhysicsB}, which become stationary when evaluated at the interacting Green's function of the system. 
Examples include applications to  atoms and molecules~\cite{Dahlen2004IntJQuantumChem,Dahlen2004JChemPhys,Dahlen2004PhysRevB,Dahlen2006PhysRevA}, to Hubbard chains~\cite{Friesen2010PhysRevB,Sabatino2021FrontChem}, or to the homogeneous electron gas~\cite{Holm2000PhysRevB,Garcia-Gonzalez2001PhysRevB,Chiarotti2022PRR}. 
When the Klein functional is combined with an optimized effective potential approach~\cite{Casida1995PhysRevA,Kummel-Kronik2008RMP} one obtains the linearized Sham-Schl\"uter equation~\cite{Sham-Schluter1983PRL,Godby-Schluter-Sham1987PRB}, which can be used to derive advanced KS-DFT functionals from diagrammatic approximations, such as the EXX+RPA exchange-correlation functional~\cite{Almbladh1999InternationalJournalofModernPhysicsB,Ismail-Beigi2010PhysRevB,Ren2012JMaterSci,Paier2012NewJPhys,Hellgren2018PhysRevB,fett-wale71book,Stefanucci-vanLeeuwen2013book,Martin-Reining-Ceperley2016book}.
Notably, the Klein functional features a term of the form $\int \frac{d\omega}{2 \pi} \text{Tr} \text{Ln} \{ G_0^{-1}G\}$ (see in Sec.~\ref{sec:framework} for more details), which is quite cumbersome to be evaluated numerically and needs dedicated treatment~\cite{Dahlen2006PhysRevA,Friesen2010PhysRevB}. The LW functional displays similar issues. 
In passing we also note that besides DFT-based and GF methods, other orbital-dependent or dynamical approaches~\cite{Marzari2021NatMater} addressing excitations are available, including dynamical mean field theory (DMFT)~\cite{Kotliar2006RevModPhys}, spectral potentials~\cite{Gatti2007PhysRevLett,Ferretti2014PhysRevB}, or Koopmans-compliant functionals~\cite{Dabo2009arXiv,Dabo2010PhysRevB,Ferretti2014PhysRevB,Nguyen2018PhysRevX}.

Importantly, dynamical potentials can be naturally employed to describe embedding situations, where the system of interest is placed in contact with an external bath. In these cases, for non-interacting systems, the embedded GF can be calculated by adding an embedding self-energy~\cite{Stefanucci-vanLeeuwen2013book,Martin-Reining-Ceperley2016book}, which has the form of a non-local and dynamical potential, to the pristine Hamiltonian. This approach has been successfully exploited, e.g., in the description of semi-infinite systems (surface Green's function) and applied to simulations of quantum transport through nanojunctions~\cite{Meir1992PhysRevLett, Haug-Jauho1996book, Nardelli1999PhysRevB,Brandbyge2002PRB,Calzolari2004PhysRevB}.
When particle interactions are considered, the situation becomes more complex, but the assumption of dealing with a non-interacting bath~\cite{Meir1992PhysRevLett} allows one to treat the problem similarly to the non-interacting case. 
Approaches such as DMFT~\cite{Kotliar2006RevModPhys}, 
which is a dynamical method targeting both total energies and spectral properties, exploit the embedding of an interacting impurity model to describe the electron-electron self-energy of strongly interacting systems.

In general, the use of dynamical potentials (e.g., originating from many-body perturbation theory~\cite{Stefanucci-vanLeeuwen2013book, Martin-Reining-Ceperley2016book}, embedding, or spectral potentials~\cite{Gatti2007PhysRevLett,Ferretti2014PhysRevB,Marzari2021NatMater}) in electronic-structure methods is a challenge by itself.
Indeed, the frequency representation of propagators (or dynamical potentials) is non trivial~\cite{Chiarotti2022PRR,Chiarotti2023arXiv,Chiarotti2023PhD} with viable approaches ranging from discretized frequency grids (both on the real or imaginary axis) to the use of meromorphic functions and Pad\'e approximants~\cite{Engel1991PhysRevB,Rojas1995PhysRevLett}, or imaginary-time treatments~\cite{Rojas1995PhysRevLett}. 
Moreover, the solution of the resulting Dyson equation (which can be cast in the form of a non-linear eigenvalue problem~\cite{Guttel2017ActaNumerica}) adds further numerical and conceptual complications 
(including multiple solutions and non-orthonormality of the eigenvectors~\cite{Golze2019FIC,Martin-Reining-Ceperley2016book,Guttel2017ActaNumerica}).
In order to address this problem,  we have  recently exploited the combination of a sum-over-poles (SOP) representation for the propagators,  with the algorithmic-inversion method (AIM)~\cite{Chiarotti2022PRR,Chiarotti2023arXiv,Chiarotti2023PhD} to exactly solve the Dyson equation resulting from dynamical potentials. 

%
%
In this work, by taking advantage of the SOP-AIM approach~\cite{Chiarotti2022PRR,Chiarotti2023arXiv,Chiarotti2023PhD},
we first derive an analytical expression for terms of the form $\text{Tr}_\omega \left\{G_0^{-1}G \right\}$, as those appearing in the Klein functional, that is valid in the general case of interacting propagators. Next, we exploit this result to ($i$) 
recover an exact expression~\cite{Ismail-Beigi2010PhysRevB} for the RPA correlation energy~\cite{{Almbladh1999InternationalJournalofModernPhysicsB,Paier2012NewJPhys,Hellgren2018PhysRevB}}, and to ($ii$)
obtain a Klein functional valid in the case of embedding where the system of interest is coupled to a non-interacting bath.

The paper is organized as follows. In Sec.~\ref{sec:framework} we present the theoretical framework used throughout the work. Next, in Sec.~\ref{sec:analytical} we derive an analytical expression for $\text{Tr}_\omega \left\{G_0^{-1}G\right\}$. Finally, in Sec.~\ref{sec:applications} we apply the newly derived result first to evaluate the RPA correlation energy, and then to the embedding of the Klein functional. Complementary details about Green's function embedding and TrLn terms are provided in Appendix~\ref{sec:GF_embedding} and Appendix~\ref{sec:complements_TrLn}, respectively.


\section{Theoretical framework}
\label{sec:framework}
\begin{figure}
\includegraphics[clip,width=0.30\textwidth]{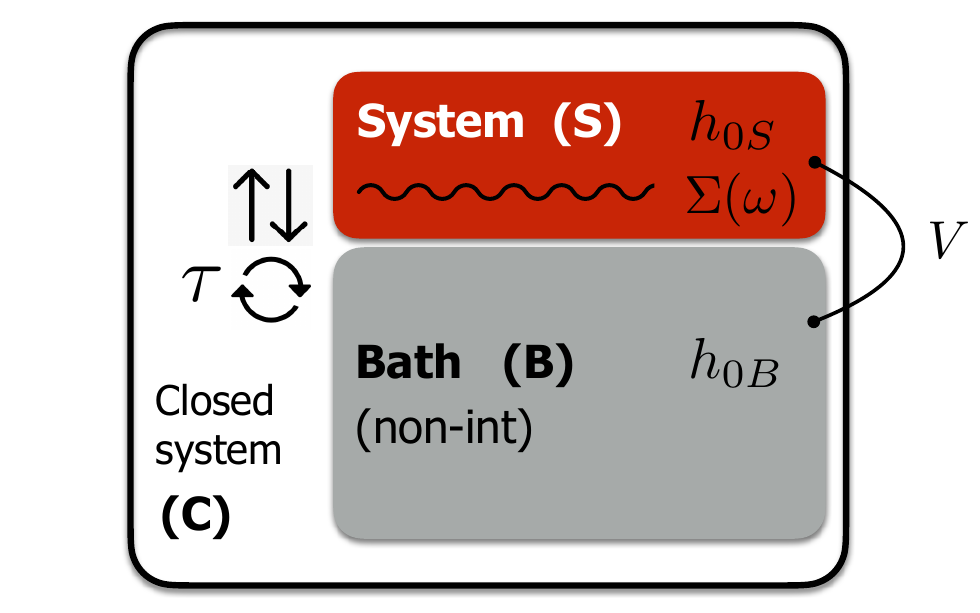}
\begin{equation}
\nonumber
\begin{array}{ccc}
   h_0,g_0  &  H_0,G_0  & H,G  \\[7pt]
   \left[ 
   \begin{array}{cc}
      h_{0S}  &   0    \\
      0       &   h_{0B}
   \end{array}
   \right]
   & 
   \left[ 
   \begin{array}{cc}
      h_{0S}  &   V   \\
      V^\dagger       &   h_{0B}
   \end{array}
   \right]   
   &
   \left[ 
   \begin{array}{cc}
      h_{0S} +\Sigma  &   V   \\
      V^\dagger       &   h_{0B}
   \end{array}
   \right]   
\end{array}
\end{equation}
\caption{Upper panel: Partitioning of the closed system $C$ into the subparts $S$ (interacting, as indicated by the wiggly line) and $B$ (non-interacting).  The Hamiltonian and self-energy blocks and the coupling $V$ of the two subsystems are also indicated.
Bottom panel: Sketch view of the three different Hamiltonians and Green's functions involved in the discussion of embedding. Left: $S$ and $B$ are non-interacting and uncoupled; Central: $S$ and $B$ are non-interacting but coupled; Right: $S$ is interacting and coupled to the non-interacting $B$. 
\label{fig:GF_embedding}}
\end{figure}

%
In this Section we present the theoretical framework underpinning the use of Green's function methods to describe an interacting system in the presence of a non-interacting bath; additional details are provided in Appendix~\ref{sec:GF_embedding}.
We consider a closed quantum system $C$ that is partitioned into two subsystems, $S$ and $B$, such that, in terms of degrees of freedom, one has $C=S \cup B$. Particle interactions are present but limited to subsystem $S$ only, leaving subsystem $B$ as a non-interacting bath. All single particle operators, including Hamiltonians, self-energies, and Green's function,  become 2$\times$2 block matrices, indexed according to the $S$ and $B$ subsystems. As detailed in Fig.~\ref{fig:GF_embedding}, $h_0$ represents the non-interacting Hamiltonian of the two systems without coupling, while $H_0$ is the non-interacting Hamiltonian of $C$ when the coupling $V$ is included. Eventually, self-energy terms accounting for the particle-particle interaction are included. As discussed in App.~\ref{sec:GF_embedding}, since interactions are only present within $S$, one can show that the corresponding self-energy is limited to the same subsystem. 
Moreover, since $h_{0B}$ is non-interacting, without loss of generality we may take it diagonal on the chosen basis, such that $h_{0B}=\text{diag}(\Omega_1,\dots,\Omega_n,\dots)$.

Within the above definitions, and following Fig.~\ref{fig:GF_embedding}, one can define the Green's functions for the whole system $C$, at different levels of description (non-interacting and uncoupled, non-interacting and coupled, interacting in $S$ and coupled), according to:
\begin{eqnarray}
   \nonumber
   g_0(\omega) &=& \left[\omega I -\text{diag}(h_{0S},h_{0B}) \right]^{-1} =   \left[\omega I -h_0 \right]^{-1} \\
   \nonumber
   G_0(\omega) &=&   \left[\omega I -H_0 \right]^{-1} \\
   G(\omega) &=& \left[\omega I -H_0 -\Sigma(\omega) \right]^{-1}
\end{eqnarray}
(time-ordered offsets from the real axis are left implicit).
We note that when $G$ is the physical GF, then $\Sigma=\Sigma_{\text{Hxc}}$ is the interaction self-energy (accounting for Hartree, exchange, and correlation terms). Nevertheless, in the following we will also consider cases where $G$ is a trial GF, as discussed, e.g., in Sec.~\ref{sec:analytical_general}.
In these cases, $\Sigma=\widetilde{\Sigma}$ just collects a set of degrees of freedom useful to represent $G$ via
\begin{equation}
  \label{eq:trial_G}
  G = G_0 + G_0 \widetilde{\Sigma} G.
\end{equation}
Within this construction, the self-energy will also be constrained to have non-zero matrix elements only within subsystem $S$, which can be seen as a domain definition for the set of trial $G$'s.

By focusing on the subsystem $S$ and making reference to the theory of Green's function embedding, the $S$ blocks of the above GFs are obtained as:
\begin{eqnarray}
   g_{0S}(\omega) &=&  \left[\omega I_S -h_{0S} \right]^{-1},  
   \nonumber \\
   G_{0S}(\omega) &=&  \left[\omega I_S -h_{0S} - \Delta v_S(\omega) \right]^{-1},  
   \nonumber \\
   G_{S}(\omega) &=&  \left[\omega I_S -h_{0S} - \Delta v_S(\omega) -\Sigma(\omega) \right]^{-1}, 
   \label{eq:embedding_GF}
\end{eqnarray}
where $\Delta v_S$ is an embedding self-energy due to the bath $B$~\cite{Haug-Jauho1996book,Stefanucci-vanLeeuwen2013book,Martin-Reining-Ceperley2016book,Nardelli1999PhysRevB,Calzolari2004PhysRevB}:
\begin{equation}
   \label{eq:embedding_SGM}
   \Delta v_S(\omega) = V g_{0B}(\omega) V^\dagger = \sum_n \frac{R_n}{\omega -\Omega_n \pm i0^+},
\end{equation}
which acts as a correction to the external potential of $S$.

The total energy of the closed system $C$ can be obtained variationally, e.g., via the Klein functional~\cite{Klein1961PR,Almbladh1999InternationalJournalofModernPhysicsB}, reading
\begin{eqnarray}
   \label{eq:Klein}
   E^K[G] &=& \text{Tr}_\omega \text{Ln} \left\{G_0^{-1} G\right\} + 
            \text{Tr}_\omega H_0G_0 \\ \nonumber
          &+&   \text{Tr}_\omega \left[ I - G_0^{-1} G\right] + \Phi_{\text{Hxc}}[G],
\end{eqnarray}
where $\Phi_{\text{Hxc}}[G]$ is a functional~\cite{Luttinger-Ward1960PR,Klein1961PR,Baym1961PhysRev} to be approximated that is related to the interaction self-energy as
\begin{equation}
   \frac{\delta \Phi_{\text{Hxc}}[G]}{\delta G} = \frac{1}{2 \pi i} \Sigma_{\text{Hxc}}[G].
\end{equation}
With the above definitions, one can show~\cite{Stefanucci-vanLeeuwen2013book,Martin-Reining-Ceperley2016book} that the gradient of the Klein functional is zero for the GF $G$ that satisfies the self-consistent Dyson equation 
\begin{equation}
  G = G_0 + G_0 \Sigma_{\text{Hxc}}[G] G.
\end{equation}

\subsection{Sum-over-poles and algorithmic inversion}
\label{sec:framework_SOP}
%
In order to make progress in the numerical exploitation of the above described techniques, in the following we make use of the concept of sum-over-poles (SOP)~~\cite{Chiarotti2022PRR,Chiarotti2023arXiv,Engel1991PhysRevB,Friesen2010PhysRevB,Sabatino2021FrontChem} to represent propagators, combined with that of the algorithmic-inversion method (AIM) to solve Dyson-like equations. 
In practice, this amounts to writing propagators and self-energies using discrete poles and residues (meromorphic representation~\cite{Friesen2010PhysRevB}) as
\begin{eqnarray}
    \label{eq:G0_SOP}
    G_0(\omega) &=& \sum_n \frac 
      {A^0_n}{\omega -\eps^0_n \pm i0^+}, 
    \\
    G(\omega)   &=& \sum_s \frac
    {A_s}{\omega -\eps_s \pm i0^+},
    \label{eq:G_SOP}
    \\
    \Sigma(\omega) &=& \Sigma_{0} + \sum_n \frac{\Gamma_{n}}{\omega -\omega_n \pm i0^+},
    \label{eq:sigma_SOP}
\end{eqnarray}
which could be seen also as discrete Lehmann representations~\cite{Engel1991PhysRevB}. 
Recently, SOPs have also been used to represent the screened Coulomb interaction in the context of GW leading to the multi-pole approximation (MPA)~\cite{Leon2021PhysRevB,Leon2023PhysRevB}.
For simplicity, in this work we assume all residues and poles to be Hermitian and real, respectively.
In the above expressions, $G_0$ is a non-interacting Green's function (GF) obtained from  the single-particle Hamiltonian $h_0$, 
\begin{eqnarray}
   h_0 | \phi^0_n\rangle = \eps^0_n | \phi^0_n\rangle, 
   \qquad A^0_n = |\phi^0_n\rangle \langle \phi^0_n |,
\end{eqnarray}
while $G$ is an interacting or embedded GF, obtained from $G_0$ by a Dyson equation involving $\Sigma$, i.e. $G=[\omega I-h_0-\Sigma(\omega)]^{-1}$. 

Having assumed discrete and real poles poles for $\Sigma$ and $G_0$ (and Hermitian residues) implies~\cite{Chiarotti2023arXiv,Guttel2017ActaNumerica} that also $G$ has real discrete poles and that the residues can be written as
\begin{eqnarray}
   \big[ h_0 + \Sigma(\epsilon_s) \big] | f_s \rangle = \epsilon_s | f_s \rangle, \qquad A_s = |f_s\rangle \langle f_s |,
\end{eqnarray}
where the normalization of $|f_s\rangle$ is defined according to
\begin{eqnarray}
   \langle f_s | f_s\rangle &=&  Z_s = 1+ 
   \langle f_s | \dot{\Sigma}(\epsilon_s) |f_s\rangle \leq 1,
   \\
   \sum_s | f_s\rangle \langle f_s | &=& I,
\end{eqnarray}
i.e., the $|f_s\rangle$ are complete though not linearly independent nor orthonormalized (see also Ref.~\cite{Chiarotti2023PhD}), where we have used $\dot{\Sigma}(\omega) = \partial \Sigma(\omega)/\partial \omega$. In writing the expressions above the Dyson equation has been mapped to a non-linear eigenvalue problem involving rational functions~\cite{Chiarotti2023arXiv,Guttel2017ActaNumerica}.
Moreover, noting that the residues of $G$ in Eq.~\eqref{eq:G_SOP} are positive semi-definite (PSD) by construction, the residues $\Gamma_{n}$ of $\Sigma$ are also forced to be PSD Hermitian operators. 
In fact, given
\begin{eqnarray}
   A(\omega) &=& \frac{1}{2 \pi i} \left[ G(\omega)-G^\dagger(\omega)\right] \text{sign}(\mu-\omega), 
   \nonumber \\
   \Gamma(\omega) &=& \frac{1}{2 \pi i} \left[ \Sigma(\omega)-\Sigma^\dagger(\omega)\right] \text{sign}(\mu-\omega),
   \\[5pt]
    A(\omega) &=& G(\omega) \Gamma(\omega) G^\dagger(\omega), 
\end{eqnarray}
(the last identity coming from the Dyson equation), the positive semi-definiteness of $A$ is equivalent~\cite{Stefanucci-vanLeeuwen2013book,
Stefanucci2014PhysRevB}
(i.e., if and only if) to that of $\Gamma$.

Next, given $G_0$ and $\Sigma$ represented as SOPs, it is possible to explicitly evaluate the coefficients of the GF $G$ solving the related Dyson equation. This approach, termed algorithmic-inversion method (AIM)~\cite{Chiarotti2023arXiv}, maps the non-linear eigenvalue problem of the Dyson equation into a linear eigen-problem in a larger space. Algebraically, this can be seen as the consequence of identifying the interaction self-energy as an embedding self-energy [see Eqs.~(\ref{eq:Gamma_factoriz}-\ref{eq:embedding_matrix})], and then solving the Hamiltonian problem in the larger subspace; details are provided in Ref.~\cite{Chiarotti2023arXiv}. We also note that similar techniques have been used in the context of dynamical mean-field theory~\cite{Savrasov2006PhysRevLett,Budich2012PhysRevB,Wang2012PhysRevB}, lattice Hamiltonians~\cite{Friesen2010PhysRevB}, and, more recently, within the GW and Bethe-Salpeter equation formalism~\cite{Bintrim2021JChemPhys,Bintrim2022JChemPhys}.
%

\section{Analytical evaluation of $\text{TrLn}$ terms}
\label{sec:analytical}
%
As a technical prerequisite for this work, and as a relevant result in itself, in this Section we focus on integrals of the form:
\begin{eqnarray}
   \nonumber
   \Delta E_K &=& \text{Tr}_\omega \text{Ln} \left\{G_0^{-1}G\right\}  \\[5pt]
       &=& \int \frac{d\omega}{2\pi i} \, e^{i\omega0^+} \, \text{Tr} \text{Ln} \left\{G_0^{-1}(\omega)G(\omega)\right\}.
       \label{eq:deltaE}
\end{eqnarray}
By representing the Green's functions $G_0$ and $G$ in the above equation as SOPs according to Eqs.~(\ref{eq:G0_SOP}-\ref{eq:G_SOP}), one can derive a general analytical expression for $\Delta E_K$ of Eq.~\eqref{eq:deltaE}, as shown below.
%
%
%
\drop{
Without loss of generality, one can introduce a self-energy operator connecting $G_0$ and $G$, by taking $\Sigma(\omega)=G_0^{-1}(\omega)-G^{-1}(\omega)$, resulting in the Dyson equation:
\begin{eqnarray}
  \label{eq:Dyson_G_Go}
  G(\omega) &=& G_0(\omega) + G_0(\omega) \Sigma(\omega) G(\omega), \\[5pt]
  \Sigma(\omega) &=& \Sigma_0 + \Delta\Sigma(\omega).
  \label{eq:Sigma_Dyson_G_Go}
\end{eqnarray}
In the last line, $\Sigma_0$ is a static operator, while $\Delta\Sigma(\omega) \to 0$ for $\omega \to \pm \infty$.
In the following we also assume $\Sigma_0$ to be Hermitian (which is physical but not necessarily granted for a given $G$).
}

In order to do this, we will make use of some common operator and matrix identities, that we report below for completeness. 
For instance, we will use the following identity:
\begin{equation}
   \text{Tr} \text{Ln}(A) = \text{Ln} \det(A).
   \label{eq:identity_TrLn}
\end{equation}
Bearing Eq.~(\ref{eq:identity_TrLn}) in mind, the following relations also hold:
\begin{eqnarray}
   \det(AB) &=& \det(A)\det(B), \\[5pt]   
   \text{Tr} \text{Ln}(AB) &=& \text{Tr} \text{Ln}(A) + \text{Tr} \text{Ln}(B).
   \label{eq:trace_ln_sum}
\end{eqnarray}
Moreover, given an operator $A$ represented in the form
\begin{equation}
  A = \left[ 
  \begin{array}{cc}
     S & V_1 \\
    V_2^\dagger & B  
  \end{array}
  \right],
\end{equation}
its determinant can be expressed according to~\cite{Brualdi1983LinearAlgebraanditsApplications}:
\begin{equation}
  \text{det}(A) = \text{det}(B)\cdot \text{det}( S- V_1B^{-1}V_2^\dagger),
  \label{eq:identity_embedding_det}
\end{equation}
which is a result reminiscent of techniques used in GF embedding, presented in Sec.~\ref{sec:framework}.

\subsection{Special case: non-interacting $G$}
\label{sec:analytical_nonint}
%
As a first step, we consider the case of both $G_0$ and $G$ in Eq.~\eqref{eq:deltaE} being non-interacting GFs corresponding to mean-field Hamiltonians $h_0$ and $h_1$, defined as:  
\begin{equation}
      h_i = \sum_m \, | \phi^i_m\rangle \eps^i_m \langle \phi^i_m | .
\end{equation}
This means that both $G_0$ and $G$ are diagonal on single-particle orthonormal basis sets ($|\phi^0_m\rangle$ and $|\phi^1_m\rangle$),  
that can be used to evaluate the traces.
Importantly, we assume that the number of occupied electrons is the same for $G_0$ and $G$.
By considering Eq.~\eqref{eq:trace_ln_sum} and 
taking $A=G_0^{-1}$ and $B=G$, one can write the $\Delta E_K$ integral as
\begin{eqnarray}
   \label{eq:det1}
   \Delta E_K &=& \int \frac{d\omega}{2 \pi i} e^{i\omega 0^+} \left[ -\text{Tr}\text{Ln} G_0 +\text{Tr}\text{Ln}G \right], \\ 
   \label{eq:det2}
       &=& \int \frac{d\omega}{2 \pi i} e^{i\omega 0^+} \Big[ 
          \text{Ln} \frac{ \Pi_m^\text{all} (\omega -\epsilon^0_m \pm i0^+)} 
                    { \Pi_m^\text{all} (\omega -\epsilon^1_m \pm i0^+)} 
     \Big].
\end{eqnarray}
The label ``all'' in the product means that both occupied and empty poles are considered.
In order to evaluate the integral using residues, the contour needs to be closed in the upper half plane, the enclosed poles corresponding to occupied states of both $G_0$ and $G$.
Since the number of occupied poles of both systems is the same, the integral $\Delta E_K$ can be re-written as
\begin{eqnarray}
   \Delta E_K &=& \sum_m^\text{occ} \,\oint_{\Gamma_m} \frac{dz}{2 \pi i}
       \text{Ln} \,\frac{z-\epsilon^0_m - i0^+}{z-\epsilon^1_m - i0^+},
       \label{eq:DE_k_contour}
\end{eqnarray}
with an example of a $\Gamma_m$ contour represented in Fig.~\ref{fig:bone_contour_integral} of App.~\ref{sec:notable_integrals}.
The analytical expression for contour integrals as those appearing in  Eq.~\eqref{eq:DE_k_contour} is provided in Eq.~\eqref{eq:bone_contour_integral} of App.~\ref{sec:notable_integrals}. Taking advantage of that expression, we recover the well-known result~\cite{Stefanucci-vanLeeuwen2013book,Martin-Reining-Ceperley2016book,Dahlen2004PhysRevB,Ismail-Beigi2010PhysRevB}: 
 \begin{equation}
   \Delta E_K = \sum_m^\text{occ} \, \left[ n^1_m\eps^1_m- n_m^0\eps^0_m \right],
 \end{equation}
where we have made the eigenvalue multiplicities $n_m^{i}$ explicit and limited the sum to distinct multiplets.

\subsection{General case: interacting $G$}
\label{sec:analytical_general}
%
Next, in this Section we consider the case of Eq.~\eqref{eq:deltaE} with a fully interacting $G$.
Without loss of generality, we can define a self-energy connecting $G$ and $G_0$ by a Dyson equation, by writing:
\begin{equation}
   \Sigma(\omega) = G_0^{-1}-G^{-1}.
\end{equation}
It is important to note that such self-energy is not necessarily physical (i.e. it may not originate from perturbation theory or from a functional formulation), but rather an auxiliary mathematical object.
Since $G_0, G$ and $\Sigma$ are connected by a Dyson equation, and having assumed discrete poles for both $G_0$ and $G$ (which then result meromorphic functions of the frequency), also $\Sigma$ has discrete poles. 
We are therefore in the condition to use the SOP representations given in Eqs.~(\ref{eq:G0_SOP}-\ref{eq:sigma_SOP}). In what follows we assume to represent single-particle operators on a truncated basis set, thereby mapping them to finite dimension matrices.

As discussed in Sec.~\ref{sec:framework_SOP}, the residues $\Gamma_n$ of $\Sigma$ are semi-positive definite (stemming from the SPD of the spectral function of $G$) and, following Refs.~\cite{Chiarotti2022PRR,Chiarotti2023arXiv}, one can introduce $V_n$ such that
\begin{equation}
   \Gamma_n = V^\dagger_n V_n.
   \label{eq:Gamma_factoriz}
\end{equation}
In doing so, $V_n$ can be taken, e.g., to be the square root of $\Gamma_n$ or to be a lower-rank rectangular matrix (when represented on a basis) if $\Gamma_n$ is low-rank. By doing this, $G$ can be seen as the GF of an embedded system (index 0, below), coupled to an external bath. Indeed, by defining the inverse resolvent ($\omega I -\mathcal{H}$) of the whole auxiliary system as
\begin{eqnarray}
       \nonumber
   \omega I-\mathcal{H} &=& \left[
   \begin{array}{c c c c c}
      \omega I -h_0 &      V_1  & V_2 & \dots & \\
             V_1^\dagger & (\omega-\omega_1)I  & &  & \\
             V_2^\dagger &  & (\omega-\omega_2)I  &  & \\
             \vdots      &  &  & \ddots & \\
   \end{array}
   \right],
   \\
   &=& \left[
   \begin{array}{c c}
      S &  V  \\
      V^\dagger & B
   \end{array}
   \right],
   \label{eq:embedding_matrix}
\end{eqnarray}
one can immediately verify that the self-energy in Eq.~\eqref{eq:sigma_SOP} is the embedding self-energy for the zeroth-block subsystem $S$ (in the following, calligraphic operators such as $\mathcal{H}$ refer to the enlarged auxiliary space).
This construction is the same used in the framework of the algorithmic inversion method~\cite{Chiarotti2022PRR,Chiarotti2023arXiv}, 
used to solve Dyson equations involving propagators represented as SOP and presented in Sec.~\ref{sec:framework_SOP}.

We can now apply the identity in Eq.~\eqref{eq:identity_embedding_det} to the matrix in Eq.~\eqref{eq:embedding_matrix}, obtaining:
\begin{eqnarray}
   \text{det}(\omega I- \mathcal{H})&=&\text{det}(B) \times \text{det}\left( S-VB^{-1}V^\dagger \right) \\
       \nonumber
       &=& \prod_{n} (\omega-\omega_n)^{r_n} \times \text{det}(\omega I-h_0-\Sigma),
\end{eqnarray}
where $r_n$ is the rank of the $\Gamma_n$ matrix.
The above equation can be recast in the following form:
\begin{eqnarray}
   \text{det}\ G(\omega) &=& \prod_{n} (\omega-\omega_n)^{r_n} \times \text{det}(\omega I- \mathcal{H})^{-1}, \\
   &=&   \frac{\prod^\text{all}_{n} (\omega-\omega_n)^{r_n}} 
         {\prod^{\text{all}}_{s} (\omega-\epsilon_s)^{n_s}},
\end{eqnarray}
where we have exploited the fact that the poles of $G$ are also eigenvalues of $\mathcal{H}$ for the whole system, and made the multiplicities $n_s$ explicit. 

Combining Eq.~\eqref{eq:det1} with the identity connecting TrLn to Ln det, Eq.~\eqref{eq:identity_TrLn}, we obtain:
\begin{eqnarray}
   \label{eq:DE_k_Ln_det}
   \Delta E_K &=& \int \frac{d\omega}{2 \pi i} e^{i\omega 0^+} \left[\text{Ln} \det G -\text{Ln} \det G_0\right], \\
   \nonumber
   &=& \int \frac{d\omega}{2 \pi i} e^{i\omega 0^+} \text{Ln}
      \left[ \frac{\prod^{\text{all}}_n (\omega-\omega_n)^{r_n} \prod^{\text{all}}_m (\omega-\epsilon^0_m)^{n_m^0}}{\prod^{\text{all}}_s (\omega-\epsilon_s)^{n_s}} \right].
   \\[7pt]
   &=& \int \frac{d\omega}{2\pi i} \, e^{i\omega0^+} \, \text{Tr} \text{Ln} \left\{\mathcal{G}_0^{-1}(\omega) \mathcal{G}(\omega)\right\}.
\end{eqnarray}
In the last equation, $\mathcal{G}$ and $\mathcal{G}_{0}$ are the GFs of the auxiliary system  obtained with and without including the coupling matrices $V$ in $\mathcal{H}$, respectively.
A counting of the degrees of freedom shows that the cardinality of $\{\epsilon_s\}$ is equal to that of $\{\epsilon^0_m\}\cup \{\omega_n\}$, as also shown by the embedding construction in Eq.~\eqref{eq:embedding_matrix}.
Nevertheless, only occupied poles (i.e. poles above the real axis) count in the integral.

If {\it the number of such poles in the numerator and in the denominator is the same}, by exploiting Eq.~\eqref{eq:DE_k_contour} we obtain the final result: 
 \begin{equation}
   \Delta E_K =  \sum_s^{\text{occ}} n_s \epsilon_s -\left[ \sum^{\text{occ}}_m n_m^0 \epsilon_m^0 +\sum^{\text{occ}}_n r_n \omega_n \right].
   \label{eq:main_result}
 \end{equation}
This expression is the first key result of the present work.
The condition of having the same number of occupied states in the numerator and denominator in the second line of Eq.~\eqref{eq:DE_k_Ln_det} is equivalent to having the same number of occupied states before and after the switch-on of the coupling matrix elements $V$. This condition, therefore, encodes charge conservation within the closed  system $C = S \cup B$.
%
%
In App.~\ref{sec:TrLn_2_interacting_GF} we also provide a generalization of Eq.~\eqref{eq:main_result} where both propagators in the TrLn term are interacting (or embedded).

\drop{
Moreover, Eq.~\eqref{eq:main_result} can be further generalized to the case of TrLn computed for two interacting GFs, $G_1$ and $G_2$. As a first step we make reference to an arbitrary non-interacting $G_0$ by exploit the identity in Eq.~\eqref{eq:trace_ln_sum},
\begin{eqnarray}
  \text{Tr}_\omega \text{Ln} \left\{G_1^{-1} G_2\right\} &=& 
  \text{Tr}_\omega \text{Ln} \left\{G_0^{-1} G_2 \right\} 
  \nonumber \\
  &-&\text{Tr}_\omega \text{Ln} \left\{G_0^{-1} G_1 \right\} .
  \label{eq:TrLn_G1G2}
\end{eqnarray}
Next we can connect $G_{1,2}$ to $G_0$ via Dyson-like equations, by writing:
\begin{eqnarray}
   G_1 &=& G_0 + G_0(\Sigma_1-v_0) G_1 \\
   G_2 &=& G_0 + G_0(\Sigma_2-v_0) G_2
\end{eqnarray}
where $\Sigma_i$ are suitable self-energy operators. Upon defining $\Sigma_{21} = \Sigma_2-\Sigma_1$, the above equations give:
\begin{equation}
  G_2 = G_1 + G_1 \Sigma_{21} G_2. 
\end{equation}
We can now evaluate Eq.~\eqref{eq:TrLn_G1G2} by means of Eq.~\eqref{eq:main_result}, obtaining:
\begin{eqnarray}
\nonumber
   \Delta E_K &=& \left[ \sum_s^{\text{occ}} n_s^{(2)}\epsilon_s^{(2)} -\sum^{\text{occ}} \text{poles}(\Sigma_2)  \right] \\
   &-& \left[ \sum_s^{\text{occ}} n_s^{(1)}\epsilon_s^{(1)} -\sum^{\text{occ}} \text{poles}(\Sigma_1) \right].
   \label{eq:main_result_gen}
\end{eqnarray}
}


At this point it is worth discussing alternative approaches existing in the literature aimed at evaluating terms of the form  $\text{Tr}_\omega \text{Ln} \left\{G_0^{-1} G_1 \right\}$.
For instance, in a series of papers, Dahlen and co-workers~\cite{Dahlen2006PhysRevA,Dahlen2004JChemPhys,Dahlen2004PhysRevB} first re-write the TrLn term of the Luttinger-Ward functional by factorizing the static part of the self-energy $\Sigma_x$, and then recasting~\cite{Dahlen2006PhysRevA} the integral for numerical integration over the imaginary axis. Along the same lines, in App.~\ref{sec:computational_eval} we provide a scheme for numerical integration of the TrLn terms that we have used in the present work to numerically validate analytical expressions such as  Eq.~\eqref{eq:main_result}.
In Ref.~\cite{Friesen2010PhysRevB}, Friesen and co-workers (which also adopt a meromorphic, i.e. SOP in our language, representation for the propagators) first handle the $\Sigma_x$ term as in Refs.~\cite{Dahlen2006PhysRevA,Dahlen2004JChemPhys,Dahlen2004PhysRevB} and then numerically evaluate the residual contribution to the integral using a coupling-constant integration.
In Ref.~\cite{Ismail-Beigi2010PhysRevB}, Ismail-Beigi discusses the RPA correlation energy in the context of Green's function theory, and, exploiting algebraic techniques similar to those employed in this work, provides an analytical expression involving the poles of the independent-particle and RPA response functions. We discuss the RPA correlation energy in Sec.~\ref{sec:applications_RPA} where we re-derive Ismail-Beigi's expression by means of the present formalism.
Additionally, Aryasetiawan et al.~\cite{Aryasetiawan2002PhysRevLett} write the RPA correlation energy in a form similar to that of Ref.~\cite{Dahlen2006PhysRevA} and App.~\ref{sec:computational_eval} for numerical evaluation along the imaginary axis.
%

\section{Applications}
\label{sec:applications}

Having derived an analytical expression for the TrLn terms defined by Eq.~\eqref{eq:deltaE}, in this Section we present two applications. First we focus on the calculation of the RPA correlation energy, providing a re-derivation of a result already known in the literature~\cite{Ismail-Beigi2010PhysRevB}, and then apply the formalism to analyze and partition the Klein functional in the presence of embedding.

\subsection{RPA correlation energy and plasmons}
\label{sec:applications_RPA}
%
\begin{figure*}
  \includegraphics[clip,width=0.7\textwidth]{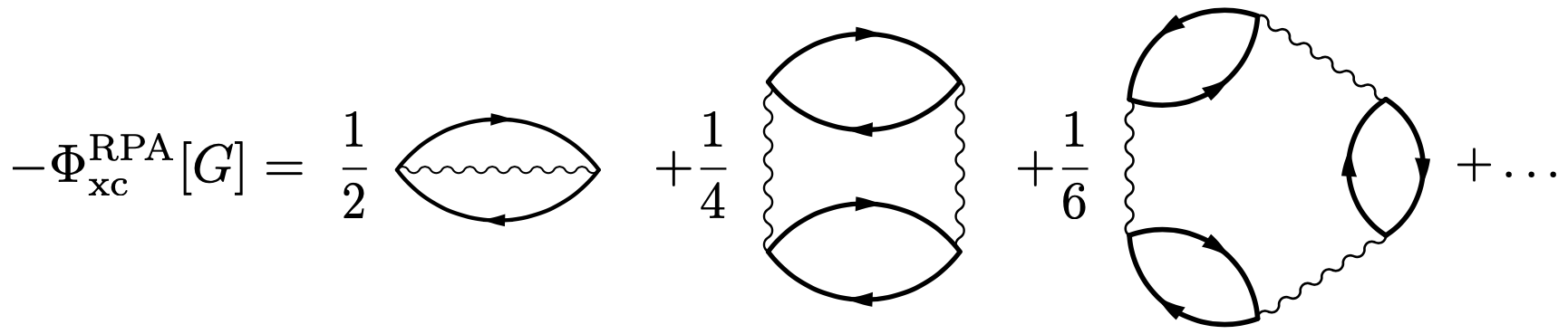}
  \caption{\label{fig:RPA} RPA exchange and correlation energy represented by means of Feynman diagrams.}
\end{figure*}

In the context of Green's function methods, 
the RPA correlation energy is written as~\cite{Almbladh1999InternationalJournalofModernPhysicsB,Ismail-Beigi2010PhysRevB,Ren2012JMaterSci,Paier2012NewJPhys,Hellgren2018PhysRevB,fett-wale71book,Stefanucci-vanLeeuwen2013book,Martin-Reining-Ceperley2016book}:
\begin{eqnarray}
\nonumber
   P(\mathbf{x}_1,\mathbf{x}_2,\omega) &=& \int \frac{d\omega'}{2\pi i}\, G(\mathbf{x}_1,\mathbf{x}_2,\omega+\omega')G(\mathbf{x}_2,\mathbf{x}_1,\omega') \\
   \Phi^{\text{RPA}}[P] &=&-\frac{1}{2}\text{Tr}_\omega \Big\{ \sum_{n=2}^\infty \frac{1}{n} \left[vP(\omega)\right]^n\Big\} \\
   &=& +\frac{1}{2}\text{Tr}_\omega \text{Ln} \big\{  I -vP(\omega) \big\} 
   \nonumber \\
   \label{eq:RPA_energy}
   & & 
   +\frac{1}{2} \text{Tr}_\omega \left\{ v P\right\},
   \\ \nonumber
   &=& \Delta\Phi^{\text{RPA}}_1 + \Delta\Phi^{\text{RPA}}_2,
\end{eqnarray}
where the irreducible polarizability $P$ is either evaluated using the Kohn-Sham Green's function $G_s$ in the optimized-effective-potential (OEP) method~\cite{Casida1995PhysRevA}, or by an interacting Green's function (e.g. at the level of self-consistent GW) when making stationary the Klein or Luttinger-Ward functionals~\cite{Klein1961PR,Luttinger-Ward1960PR,Baym1961PhysRev,Stefanucci-vanLeeuwen2013book,Martin-Reining-Ceperley2016book}. By considering the Dyson equation
\begin{equation}
  \chi(\omega) = P(\omega) + P(\omega) v\chi(\omega),
  \label{eq:dyson_polarizability}
\end{equation}
connecting the irreducible and reducible polarizabilities ($P$ and $\chi$, respectively), one obtains 
\begin{equation}
   I -vP = \epsilon = \chi^{-1}P,
\end{equation}
which can be used in the first term $\Delta\Phi^{\text{RPA}}_1$of Eq.~\eqref{eq:RPA_energy}, leading to:
\begin{equation}
  \Phi^{\text{RPA}}[P] = -\frac{1}{2}\text{Tr}_\omega \text{Ln} \big\{ \chi P^{-1}\big\} 
  +\frac{1}{2} \text{Tr}_\omega \left\{ v P\right\}.
\end{equation}
By considering the $\chi$ and $P$ as two interacting single particle propagators, we can apply Eqs.~(\ref{eq:Dyson_simga21}-\ref{eq:main_result_gen}) with $\Sigma_{21}=v$ in view of Eq.~\eqref{eq:dyson_polarizability}. This means that the poles of the two self-energies need to cancel out identically and therefore do not contribute to the evaluation of the Tr Ln term.
In turn, we obtain:
\begin{eqnarray}
   \Delta\Phi^{\text{RPA}}_1 &=& -\frac{1}{2} 
   \sum_{p}^{\text{occ}} \left[ n_p\Omega_p -n_p^0 \Omega_p^{0}\right],
   \nonumber
   \\
   &=& \phantom{+}\frac{1}{2} \sum_{p}^{\Omega_p>0} 
   \left[ n_p \Omega_p - n_p^0\Omega_p^{0}\right],
\end{eqnarray}
where $\Omega_p$ and $\Omega^0_p$ are the poles of $\chi$ and $P$ respectively, and we have considered that each time-ordered polarizability has poles at $\pm|\Omega^{(0)}_p|$, the negative ones being those above the real axis and contributing to the integral. Degeneracies of the poles ($n_p$ and $n_p^0$), have been marked explicitly.

We now turn to the evaluation of the second term, $\Delta\Phi^{\text{RPA}}_2$ in Eq.~\eqref{eq:RPA_energy}.
The irreducible polarizability $P$ can be represented as a sum-over-poles according to:
\begin{equation}
  P(\omega) = \sum_p^{\Omega_p>0} \left[ \frac{|t_p\rangle\langle t_p|} 
  {\omega -\Omega^{0}_p +i0^+} -\frac{|t_p\rangle\langle t_p|}{\omega +\Omega^{0}_p -i0^+}
  \right],
\end{equation}
where $\langle \mathbf{x}|t\rangle = \phi_c(\mathbf{x})\phi_v^*(\mathbf{x})$, $(c,v)$ referring to conduction and valence single particle orbitals, respectively. 
With the above definitions, one obtains:
\begin{eqnarray}
   \Delta\Phi^{\text{RPA}}_2 = -\frac{1}{2} \sum_p^{\Omega_p>0} \langle t_p|v|t_p\rangle,
\end{eqnarray}
which completes the evaluation of the RPA correlation energy, consistently with existing literature. In particular, we have recovered Eq.~(23) of Ref.~\cite{Ismail-Beigi2010PhysRevB}.

\subsection{Embedding of the Klein functional}
\label{sec:applications_klein}
%
The main goal of the present Section is to study the Klein functional in the presence of an embedding scheme as the one described in Sec.~\ref{sec:framework} and App.~\ref{sec:GF_embedding}, in order to derive, as demonstrated below, a variational partition of the total energy.
In order to do so we begin by partitioning each term appearing in the Klein functional given by Eq.~\eqref{eq:Klein}.
Notably, the functional depends on a trial Green's function $G$ that, according to Eq.~\eqref{eq:trial_G}, we represent by means of a self-energy $\widetilde{\Sigma}$ constrained to be localized on the subsystem $S$. As discussed in Sec.~\ref{sec:framework}, this represents a definition for the domain of the trial GF $G$.

For what concern $\Phi_{\text{Hxc}}$, the partition is already in place since the particle-particle interaction is only present in $S$. Therefore one has
\begin{equation}
  \Phi_{\text{Hxc}}[G] = \Phi_{\text{Hxc}}[G_S].
\end{equation}
This can be understood, e.g., diagrammatically, since the bare interaction lines only connect points in the $S$ subsystem, making each vertex located in $S$. This is further discussed in App.~\ref{sec:GF_embedding}.
Next we consider the $\text{Tr}_\omega H_0G_0$ term, which is the non-interacting energy of the closed $C=S\cup B$ system, and can be partitioned as
\begin{eqnarray}
   \text{Tr}_\omega \left\{ H_0G_0 \right\} &=& 
   \nonumber
      \text{Tr}^S_\omega \left\{ (h_{0S}+\Delta v_S)G_{0S} \right\} \\
      &+& 
      \text{Tr}^B_\omega \left\{ (h_{0B}+\Delta v_B)G_{0B} \right\} \\
      &=& \sum_s^{\text{occ}} \epsilon_s^0,
\end{eqnarray}
where $\epsilon_s^0$ are the eigenvalues of the non-interacting problem for $C$, $H_0 |\phi_s\rangle = \epsilon_s^0 | \phi_s\rangle$.

Coming to the next term, the following chain of identities also holds
\begin{eqnarray}
   \text{Tr}_\omega \left\{ I-G_0^{-1}G \right\} &=& -\text{Tr}_\omega \left\{ \widetilde{\Sigma}G \right\}  \nonumber \\
     &=& -\text{Tr}^S_\omega \left\{ \widetilde{\Sigma}_SG_S \right\}, \nonumber \\
     &=& \phantom{+}\text{Tr}^s_\omega \left\{ I_S-G_{0S}^{-1}G_S \right\},
\end{eqnarray}
where we have represented the trial $G$ according to Eq.~\eqref{eq:trial_G}, and limiting $\widetilde{\Sigma}$ to have non-zero matrix elements only in $S$ and to have a regular propagator-like analytical structure featuring time-ordering and simple (first order) poles. Indeed, the last step is valid because of the following equation:
\begin{equation}
  G_S = G_{0S} + G_{0S} \widetilde{\Sigma}_SG_S .
\end{equation}

The last and most interesting term in Eq.~\eqref{eq:Klein} is  $\text{Tr}_\omega \text{Ln} G_0^{-1}G$, which can be evaluated using Eq.~\eqref{eq:main_result}:
\begin{eqnarray}
   \nonumber 
    \text{Tr}_\omega \text{Ln} \left\{G_0^{-1} G\right\} &=& \sum^{\text{occ}}_s \epsilon_s -\sum_n^{\text{occ}} \epsilon^0_n 
    -\sum_n^{\text{occ}} \text{poles}(\widetilde{\Sigma})
    \\
    &=&\text{Tr}^S_\omega \text{Ln} \left\{G_{0S}^{-1} G_S\right\},
\end{eqnarray}
where we have used the fact that $\sum_s \epsilon_s = \sum \text{poles}(G_{S})$, $\sum_n \epsilon^0_n = \sum \text{poles}(G_{0S})$. 
Using the notation introduced in Eqs.~(\ref{eq:embedding_GF}-\ref{eq:embedding_SGM}) where $\Omega_n$ are the poles of the embedding self-energy, one can show that
the term $\sum_n \Omega_n = \sum \text{poles}(\Delta v_S)$ does not explicitly appear because the embedding self-energy is used in the evaluation of both the $G_{0S}$ and $G_{S}$ Green's functions. Multiplicities have been kept implicit in the sums over eigenvalues.

Alternatively, the same result can be obtained directly from the use of Eq.~\eqref{eq:identity_TrLn} and the identity concerning the determinant of block matrices, Eq.~\eqref{eq:identity_embedding_det}.
In particular, from
\begin{equation}
G^{-1}(\omega) =\left[
  \begin{array}{cc}
      \omega I_S -h_{0S}-\widetilde{\Sigma} &
      -V  \\
      -V^\dagger  & \omega I_B -h_{0B}
  \end{array}
\right]
\end{equation}
one gets
\begin{eqnarray}
  \text{det}G^{-1} &=&  \text{det}g^{-1}_{0B} \times \text{det} G_S^{-1}, \\
  \text{det}G^{-1}_0 &=&  \text{det}g^{-1}_{0B} \times \text{det} G_{0S}^{-1},
\end{eqnarray}
which gives
\begin{eqnarray}
  \text{Tr}_\omega \left\{ G_0^{-1}G\right\} &=&
       -\text{Ln det} g^{-1}_{0B} -\text{Ln det} G^{-1}_{S} \\ \nonumber
       & &+ \text{Ln det} g^{-1}_{0B} +\text{Ln det} G^{-1}_{0S} \\ \nonumber
       &=& \phantom{+}\text{Ln det} G^{-1}_{0S} G_{S},
\end{eqnarray}
the last line being equivalent to the result to be proven.

We are now in the position to put all terms together to obtain:
\begin{eqnarray}
 E_K[G] &=& \text{Tr}^S_\omega \text{Ln} \left\{ G_S G^{-1}_{0S}\right\}
       + \sum_s^{\text{occ}} \epsilon_s^0
 \\ \nonumber
       &+& \text{Tr}^S_\omega \left\{I_S -G_{0S}^{-1}G_S \right\} + \Phi_{\text{Hxc}}[G_S].
\end{eqnarray}
Next, the first term on the rhs can be further rewritten using:
\begin{eqnarray}
   \nonumber
    \text{Tr}^S_\omega \text{Ln} \left\{ G_S G^{-1}_{0S}\right\} &=&  \text{Tr}^S_\omega \text{Ln} \left\{ G_S g_{0S}^{-1} \, g_{0S}G^{-1}_{0S}\right\}
    \\ \nonumber
    &=& 
         \text{Tr}^S_\omega \text{Ln} \left\{ G_S g^{-1}_{0S}\right\} 
         \\ 
         &-& \text{Tr}^S_\omega \text{Ln} \left\{ G_{0S} g^{-1}_{0S}\right\},
         \\[5pt]
     \text{Tr}^S_\omega \text{Ln} \left\{ G_{0S} g^{-1}_{0S}\right\} &=& \sum_s^{\text{occ}} \epsilon_s^0 - \sum_s^{\text{occ}} \bar{\epsilon}_s^0 -\sum_n^{\text{occ}} \Omega_n   \\
     &=& \sum_s^{\text{occ}} \epsilon_s^0
     -\text{Tr}^S_\omega \left\{h_{0S}g_{0S}\right\} 
     \nonumber \\
     &-&\text{Tr}^B_\omega \left\{h_{0B}g_{0B}\right\},
\end{eqnarray}
where the eigenvalues $\bar{\epsilon}_s^0$ refer to susbsystem $S$ in the absence of coupling to $B$.

Eventually, this leads to the final result for the partitioning of the Klein energy functional:
\begin{eqnarray}
  \label{eq:klein_embedding_tot}
 E_K[G] &=&  E^S_K[G_S] + \text{Tr}^B_\omega \left\{h_{0B}g_{0B}\right\},
 \\[5pt]
\nonumber
 E^S_K[G_S] &=& \text{Tr}^S_\omega \text{Ln} \left\{ G_S g^{-1}_{0S}\right\} + \text{Tr}^S_\omega \left\{h_{0S}g_{0S}\right\} 
 \\[3pt] \nonumber
       &+& \text{Tr}^S_\omega \left\{I_S -g_{0S}^{-1}G_S \right\} + 
           \text{Tr}^S_\omega \left\{\Delta v_S G_S \right\}
 \\[3pt]      
       &+& \Phi_{\text{Hxc}}[G_S] .
       \label{eq:klein_embedding}
\end{eqnarray}
This is the second key result of the present paper, implying that $E^S_K[G_S]$ is stationary for the $G_S$ that solve the embedding Dyson equation, namely:
\begin{equation}
  2\pi i\frac{\delta E^S_K[G_S]}{\delta G_S} = G_S^{-1} -g_{0S}^{-1} + \Delta v_S + \Sigma_{\text{Hxc}}[G_S] = 0, 
\end{equation}
showing that the partition of the Klein energy is exact and also variational for what concern subsystem $S$.

Interestingly, we note that an equation formally equivalent to Eq.~\eqref{eq:klein_embedding} has been used by Savrasov and Kotliar in Refs.~\cite{Savrasov2004PhysRevB,Kotliar2006RevModPhys} to express the grand-potential of a quantum system in the presence of an external local and dynamical potential coupled to the local Green's function. In the present context, that term is played by $\Delta v_S$, here originating from an embedding procedure.
Interestingly, the embedding construction allows us to further inspect the physical nature of the energy terms in Eqs.~(\ref{eq:klein_embedding_tot}-\ref{eq:klein_embedding}). In particular, the complement energy $\text{Tr}^B_\omega \left\{h_{0B}g_{0B}\right\}$ (i.e. the energy that needs to be summed to $E^S_K[G_S]$ to give the total energy of the closed system $C$, $E_K[G]$) is that of the non-interacting and uncoupled bath. This means that all effects of the coupling $V$ need to be absorbed in $E^S_K[G_S]$ to allow for variationality. This is at variance with other possible partitions of the $C$ total energy (such as, e.g., those suggested by the Galitskii-Migdal expression).

%

\section{Conclusions}
%
In this work, and within the framework of Green's function methods, we address the use of the Klein functional when embedding an interacting system $S$ into a non-interacting bath $B$.
Exploiting a sum-over-pole (SOP) representation for the propagators, and taking advantage of the algorithmic-inversion method (AIM) introduced to solve Dyson-like equations involving SOP propagators~\cite{Chiarotti2022PRR,Chiarotti2023arXiv}, we have first derived an exact analytical expression to evaluate terms of the form $\text{Tr}_\omega \text{Ln} \left\{ G_0^{-1}G\right\}$.
Notably, such terms appear in the Klein and Luttinger-Ward functionals~\cite{Klein1961PR,Luttinger-Ward1960PR,Baym1961PhysRev,Stefanucci-vanLeeuwen2013book,Martin-Reining-Ceperley2016book} as well as in other common maby-body terms such as the RPA correlation energy~\cite{Almbladh1999InternationalJournalofModernPhysicsB,Ismail-Beigi2010PhysRevB,Ren2012JMaterSci,Paier2012NewJPhys,Hellgren2018PhysRevB,fett-wale71book,Stefanucci-vanLeeuwen2013book,Martin-Reining-Ceperley2016book}. In this respect, the analytical expression obtained represents the first key result of the paper.

Next, we have used the above analytical result to partition the Klein functional of an embedded system intro two contributions, one associated to the subsystem $S$ and one to the non-interacting bath $B$. Importantly, the energy associated to $S$ is also variational as a functional of the $S$ Green's function $G_S$, with the functional gradient becoming zero for the physical embedded $G_s$. This is the second main result of the work.
Last, we have also exploited the analytical result for the TrLn terms to recover an exact analytical expression for the RPA correlation energy known in the literature~\cite{Ismail-Beigi2010PhysRevB}.

\section{Acknowledgments}
We thank Prof. Marco Gibertini and Prof. Lucia Reining for useful discussions on the subject. We also thank Matteo Quinzi for reading the manuscript and for providing further numerical validation for some of the analytical results presented.
%

%
\appendix
%
%

\section{Green's function embedding and perturbation theory}
\label{sec:GF_embedding}
%
In this Appendix we discuss the building of many-body perturbation theory (MBPT), to include particle interaction effects in the Green's function in the presence of embedding. We consider the case of fermions at $T=0$, for simplicity.
As mentioned in Sec.~\ref{sec:framework} and sketched in Fig.~\ref{fig:GF_embedding},
we consider a closed quantum system $C$ partitioned into two sub-units, $C=S\cup B$, interacting
via a coupling potential $V$, with particle interactions confined to the $S$ region, with $B$ being a non-interacting bath. 
The particle-particle interaction can be written in the usual form of a two-body potential:
\begin{eqnarray}
  V_{ee} &=& \frac{1}{2}\int d\mathbf{x}d\mathbf{x}' \, \hat{\psi}^\dagger(\mathbf{x})\hat{\psi}^\dagger(\mathbf{x}') \,
                   v_{\text{int}}(\mathbf{x},\mathbf{x}') \, \hat{\psi}(\mathbf{x}')\hat{\psi}(\mathbf{x}),
  \nonumber
  \\[5pt] 
  & &v_{\text{int}}(\mathbf{x},\mathbf{x}') \neq 0 \qquad \text{for} \quad \mathbf{x},\mathbf{x}' \in S,
  \label{eq:S_interaction}
\end{eqnarray}
where the constraint on $v_{\text{int}}(\mathbf{x},\mathbf{x}')$ expresses the fact that the interaction is present only in the $S$ region.

Within the above definitions, the perturbation expansion for the Green's function of the closed system $C$ leads to~\cite{fett-wale71book,Stefanucci-vanLeeuwen2013book,Martin-Reining-Ceperley2016book}:
\begin{multline}
  \label{eq:MBPT_G}
  iG(\mathbf{x},t;\mathbf{x}',t') = 
        \sum_{n=0}^\infty\,\frac{(-i)^n}{ n!} 
        \int_{-\infty}^{+\infty} dt_1 \dots
        dt_n 
        \\ 
        \times
        \frac{\langle \Phi_0 |\mathcal{T}\Big[\hat{V}_{ee}(t_1)\dots \hat{V}_{ee}(t_n)\, 
              \hat{\psi}(\mathbf{x},t)\hat{\psi}^\dagger(\mathbf{x}',t')\Big]| \Phi_0\rangle}
             {\langle \Phi_0|\hat{S}|\Phi_0\rangle}, 
\end{multline}
\\[-30pt]
\begin{multline}
  \label{eq:MBPT_S}
   {\langle \Phi_0|\hat{S}|\Phi_0\rangle}
     = 
        \sum_{n=0}^\infty\,\frac{(-i)^n}{ n!} 
        \int_{-\infty}^{+\infty} dt_1 \dots
        dt_n 
        \\ 
        \times
        \langle \Phi_0 |\mathcal{T}\Big[\hat{V}_{ee}(t_1)\dots \hat{V}_{ee}(t_n)\, 
        \Big]| \Phi_0\rangle .
\end{multline}
First we focus on $G^S$, i.e., on the case when $\mathbf{x},\mathbf{x}'$ are located in $S$. 
Since $\hat{V}_{ee}$ only contains field operators related to subspace $S$, all self-energy diagrams 
resulting from Eq.~(\ref{eq:MBPT_G}) have only vertexes within the subsystem $S$.
Similarly, if we consider $G$ in the general case (end points either in $B$ or $S$), $B$ points will be present only in disconnected diagrams (to be dropped) or in the external ends of the connected diagrams, which do not show in the proper self-energy. Therefore, the interaction self-energy is zero for matrix elements out of the $S$ block, as shown in 
Fig.~\ref{fig:GF_embedding}.

So far, perturbation theory in terms of the bare Green's function $G_0$ has been addressed, with $\Sigma^S[G_0]=\Sigma^S[G_0^S]$. Nevertheless, 
one can perform the usual steps~\cite{fett-wale71book,Stefanucci-vanLeeuwen2013book,Martin-Reining-Ceperley2016book} in passing from bare diagrams involving $G_0$ to skeleton diagrams involving $G$, leading to:
\begin{equation}
  \Sigma^S[G]=\Sigma^S[G^S],
\end{equation}
where we can substitute $G^S$ to $G$ because of the localization of the bare interaction, Eq.~\eqref{eq:S_interaction}.
A similar reasoning can be applied to the $\Phi$ functional to obtain $\Phi_{\text{Hxc}}[G]=\Phi_{\text{Hxc}}[G^S]$.
In summary, within the non-interacting bath condition, the interaction self-energy $\Sigma^S$ 
has a perturbation expansion structurally identical to the one usually developed for closed systems~\cite{fett-wale71book,Stefanucci-vanLeeuwen2013book,Martin-Reining-Ceperley2016book}, 
and does not make any reference to the $B$ unit, i.e. all diagrams 
develop within $S$, as if $S$ were disconnected from $B$. 
Of course, $G^S$ is then calculated in the presence of the bath, e.g. via embedding self-energies, which in turn make the effect of the interaction spread all over the system.
Notably, the Anderson impurity model~\cite{Anderson1961PhysRev,Kotliar2006RevModPhys,Martin-Reining-Ceperley2016book} can be seen as a special case of the above setting. Indeed, the exact electron-electron self-energy of the model is localized on the impurity~\cite{Anderson1961PhysRev} ($S$ in our notation), and can be computed, e.g., using bare perturbation theory~\cite{Yosida1970ProgressofTheoreticalPhysicsSupplement, Yamada1975ProgressofTheoreticalPhysics,Horvatic1987PhysRevB,Kotliar2006RevModPhys} involving $G^S_0$.

As a relevant point for the present discussion, the use of the skeleton perturbation theory and the Luttinger-Ward functional has been recently questioned~\cite{Kozik2015PhysRevLett,Eder2014arXiv,Stan2015NewJPhys,Rossi2015JPhysA:MathTheor,Schafer2016PhysRevB,Tarantino2017PhysRevB}, leading to a discussion about the domain of the trial $G$ and the rise of multiple solutions of the non-linear Dyson equation involving $\Sigma[G]$ (see e.g. Ref.~[\onlinecite{Tarantino2017PhysRevB}] for additional details). 
For the sake of the present work, we assume to be in the situation where perturbation theory does not pose convergence problems and one is able to discriminate between physical from unphysical solutions when needed.
%
%

\section{Complements on TrLn terms}
\label{sec:complements_TrLn}

\subsection{Notable integrals}
\label{sec:notable_integrals}
\begin{figure}
   \includegraphics[clip,width=0.35\textwidth]{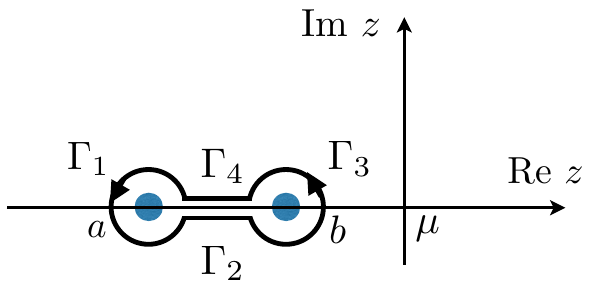}
   \caption{\label{fig:bone_contour_integral} Illustration of the contour used in Eq.~\eqref{eq:bone_contour_integral} and its decomposition in simple paths, $\Gamma_1-\Gamma_4$.}
\end{figure}
In this Section we provide a detailed derivation of the expression
\begin{equation}
   I = \oint_{\Gamma} \frac{dz}{2 \pi i} \,
       \text{Ln} \,\frac{z-a}{z-b} = b-a,
   \label{eq:bone_contour_integral}
\end{equation}
where both $a,b$ are assumed to be real numbers.
Making reference to Fig.~\ref{fig:bone_contour_integral}, the contour integral can be split into four contributions, labelled $\Gamma_1-\Gamma_4$, such that $I=I_1+I_2+I_3+I_4$, with $I_i = \int_{\Gamma_i}[...]$. 

Let us first consider $I_1$, where we assume that $\Gamma_1$ corresponds to the pole in $a$. Using the parametrization $z=Re^{i\theta}$ one has:
\begin{eqnarray}
  I_1 &=& 
  \int_{\Gamma_1} \frac{dz}{2 \pi i} \text{Ln}\frac{z-a}{z-b}, 
      \nonumber \\
      &=& 
      -R \int_{0}^{2\pi} \frac{d\theta}{2 \pi} e^{i\theta} \text{Ln}\left[1+\frac{a-b}{R e^{i\theta}}\right],
\end{eqnarray}
which goes to zero in the limit $R\to 0$, e.g. in view of $R \text{ln}(1/R)\to 0$.
A similar argument holds for $I_3$, so that we have
   $I_{1,3} \to 0$ when $R\to 0$.
Coming to remaining paths, we have
\begin{eqnarray}
  I_{2+4} = \frac{1}{2 \pi i}\left[ -\int_{a+R}^{b-R}\!\!\!dz^+ + \int_{a+R}^{b-R}\!\!\!dz^- \right] \,\text{Ln}\frac{z-a}{z-b},
\end{eqnarray}
where $dz^+$ and $dz^-$ refer to the upper ($\Gamma_4$) and lower ($\Gamma_2$) branch, respectively. The real part of the logarithm function does not contribute (the two branches cancel out), while the imaginary part does. Indeed, choosing the branch cut of the complex Log going from 0 to $+\infty$, one obtains:
\drop{
\begin{eqnarray}
   \text{Im}\, \text{Ln}(z-a) &=& \left\{
   \begin{array}{cl}
      0,       & \qquad {\text{upper br.}} \\
      2 \pi i, & \qquad {\text{lower br.}} 
   \end{array}
   \right.  \\
   \text{Im}\, \text{Ln}(z-b) &=& \left\{
   \begin{array}{cl}
      \pi i,  & \qquad {\text{upper br.}} \\
      \pi i,  & \qquad {\text{lower br.}} 
   \end{array}
   \right. 
\end{eqnarray}
This results in 
}
\begin{equation}
  I_{2+4} = \frac{1}{2 \pi}(\pi -0 + 2\pi-\pi)(b-a)=b-a ,
\end{equation}
which completes the derivation of Eq.~\eqref{eq:bone_contour_integral}.

\subsection{Computational evaluation of TrLn terms}
\label{sec:computational_eval}
%
In order to develop a form of Eq.~\eqref{eq:deltaE} suitable for numerical evaluation, that we have used e.g. to compare with the analytical results of this work, we follow some of the ideas from the App.~B of Ref.~[\onlinecite{Dahlen2006PhysRevA}]. We start by re-writing Eq.~\eqref{eq:DE_k_Ln_det} by rotating the integration over the imaginary axis:
\begin{eqnarray}
   \Delta E_K &=& 
   \int_{+i\infty}^{-i\infty} \frac{d z}{2 \pi i} \text{Ln} \left[ \frac{\det G(z)}{\det G_0(z)}\right], 
   \\
   &=& 
   \int_{-\infty}^{+\infty} \frac{d x}{2 \pi} \text{Ln} \left[\frac{\det G(ix)}{\det G_0(ix)}\right], 
   \nonumber 
   \\[8pt]
   &=&\int_{0}^{+\infty} \frac{d x}{2 \pi} \bigg[\text{Ln} \det G(ix) + \text{Ln}\, {\det}^*G(ix) \nonumber \\
   & &\quad\quad\quad\ \  -\text{Ln}\, {\det} G_0(ix) -\text{Ln}\, {\det}^* G_0(ix)\bigg],
   \nonumber
   \\[7pt]
   &=&\int_{0}^{+\infty} \frac{d x}{2 \pi} \bigg[ \,\text{ln} {\big|\det G(ix)\big|^2} -\text{ln}{\big|{\det} G_0(ix)\big|^2} \bigg].
   \nonumber \\
\end{eqnarray}
In deriving these equations we have made use of the relations $G(-ix)=G(ix)^\dagger$ and $\det M^\dagger = (\det M)^*$.
The last expression is suited for numerical evaluation, that we performed using a tangent grid on the imaginary axis.

\subsection{TrLn term with two interacting Green's functions}
\label{sec:TrLn_2_interacting_GF}
%
As anticipated in Sec.~\ref{sec:analytical_general}, Eq.~\eqref{eq:main_result} can be further generalized to the case of TrLn computed for two interacting GFs, $G_1$ and $G_2$. As a first step we make reference to an arbitrary non-interacting $G_0$ by exploit the identity in Eq.~\eqref{eq:trace_ln_sum},
\begin{eqnarray}
  \text{Tr}_\omega \text{Ln} \left\{G_1^{-1} G_2\right\} &=& 
  \text{Tr}_\omega \text{Ln} \left\{G_0^{-1} G_2 \right\} 
  \nonumber \\
  &-&\text{Tr}_\omega \text{Ln} \left\{G_0^{-1} G_1 \right\} .
  \label{eq:TrLn_G1G2}
\end{eqnarray}
Next we can connect $G_{1,2}$ to $G_0$ via Dyson-like equations, by writing:
\begin{eqnarray}
   G_1 &=& G_0 + G_0(\Sigma_1-v_0) G_1 \\
   G_2 &=& G_0 + G_0(\Sigma_2-v_0) G_2
\end{eqnarray}
where $\Sigma_i$ are suitable self-energy operators. Upon defining $\Sigma_{21} = \Sigma_2-\Sigma_1$, the above equations give:
\begin{equation}
  G_2 = G_1 + G_1 \Sigma_{21} G_2. 
  \label{eq:Dyson_simga21}
\end{equation}
We can now evaluate Eq.~\eqref{eq:TrLn_G1G2} by means of Eq.~\eqref{eq:main_result}, obtaining:
\begin{eqnarray}
\nonumber
   \Delta E_K &=& \left[ \sum_s^{\text{occ}} n_s^{(2)}\epsilon_s^{(2)} -\sum^{\text{occ}} \text{poles}(\Sigma_2)  \right] \\
   &-& \left[ \sum_s^{\text{occ}} n_s^{(1)}\epsilon_s^{(1)} -\sum^{\text{occ}} \text{poles}(\Sigma_1) \right].
   \label{eq:main_result_gen}
\end{eqnarray}


\begin{thebibliography}{75}%
\makeatletter
\providecommand \@ifxundefined [1]{%
 \@ifx{#1\undefined}
}%
\providecommand \@ifnum [1]{%
 \ifnum #1\expandafter \@firstoftwo
 \else \expandafter \@secondoftwo
 \fi
}%
\providecommand \@ifx [1]{%
 \ifx #1\expandafter \@firstoftwo
 \else \expandafter \@secondoftwo
 \fi
}%
\providecommand \natexlab [1]{#1}%
\providecommand \enquote  [1]{``#1''}%
\providecommand \bibnamefont  [1]{#1}%
\providecommand \bibfnamefont [1]{#1}%
\providecommand \citenamefont [1]{#1}%
\providecommand \href@noop [0]{\@secondoftwo}%
\providecommand \href [0]{\begingroup \@sanitize@url \@href}%
\providecommand \@href[1]{\@@startlink{#1}\@@href}%
\providecommand \@@href[1]{\endgroup#1\@@endlink}%
\providecommand \@sanitize@url [0]{\catcode `\\12\catcode `\$12\catcode
  `\&12\catcode `\#12\catcode `\^12\catcode `\_12\catcode `\%12\relax}%
\providecommand \@@startlink[1]{}%
\providecommand \@@endlink[0]{}%
\providecommand \url  [0]{\begingroup\@sanitize@url \@url }%
\providecommand \@url [1]{\endgroup\@href {#1}{\urlprefix }}%
\providecommand \urlprefix  [0]{URL }%
\providecommand \Eprint [0]{\href }%
\providecommand \doibase [0]{http://dx.doi.org/}%
\providecommand \selectlanguage [0]{\@gobble}%
\providecommand \bibinfo  [0]{\@secondoftwo}%
\providecommand \bibfield  [0]{\@secondoftwo}%
\providecommand \translation [1]{[#1]}%
\providecommand \BibitemOpen [0]{}%
\providecommand \bibitemStop [0]{}%
\providecommand \bibitemNoStop [0]{.\EOS\space}%
\providecommand \EOS [0]{\spacefactor3000\relax}%
\providecommand \BibitemShut  [1]{\csname bibitem#1\endcsname}%
\let\auto@bib@innerbib\@empty
\bibitem [{\citenamefont {Kohn}(1999)}]{Kohn1999RevModPhys}%
  \BibitemOpen
  \bibfield  {author} {\bibinfo {author} {\bibfnamefont {W.}~\bibnamefont
  {Kohn}},\ }\href {\doibase 10.1103/RevModPhys.71.1253} {\bibfield  {journal}
  {\bibinfo  {journal} {Rev. Mod. Phys.}\ }\textbf {\bibinfo {volume} {71}},\
  \bibinfo {pages} {1253} (\bibinfo {year} {1999})}\BibitemShut {NoStop}%
\bibitem [{\citenamefont {Pribram-Jones}\ \emph {et~al.}(2015)\citenamefont
  {Pribram-Jones}, \citenamefont {Gross},\ and\ \citenamefont
  {Burke}}]{Pribram-Jones2015AnnuRevPhysChem}%
  \BibitemOpen
  \bibfield  {author} {\bibinfo {author} {\bibfnamefont {A.}~\bibnamefont
  {Pribram-Jones}}, \bibinfo {author} {\bibfnamefont {D.~A.}\ \bibnamefont
  {Gross}}, \ and\ \bibinfo {author} {\bibfnamefont {K.}~\bibnamefont
  {Burke}},\ }\href {\doibase 10.1146/annurev-physchem-040214-121420}
  {\bibfield  {journal} {\bibinfo  {journal} {Annu. Rev. Phys. Chem.}\ }\textbf
  {\bibinfo {volume} {66}},\ \bibinfo {pages} {283–304} (\bibinfo {year}
  {2015})}\BibitemShut {NoStop}%
\bibitem [{\citenamefont {Marzari}\ \emph {et~al.}(2021)\citenamefont
  {Marzari}, \citenamefont {Ferretti},\ and\ \citenamefont
  {Wolverton}}]{Marzari2021NatMater}%
  \BibitemOpen
  \bibfield  {author} {\bibinfo {author} {\bibfnamefont {N.}~\bibnamefont
  {Marzari}}, \bibinfo {author} {\bibfnamefont {A.}~\bibnamefont {Ferretti}}, \
  and\ \bibinfo {author} {\bibfnamefont {C.}~\bibnamefont {Wolverton}},\ }\href
  {\doibase 10.1038/s41563-021-01013-3} {\bibfield  {journal} {\bibinfo
  {journal} {Nat. Mater.}\ }\textbf {\bibinfo {volume} {20}},\ \bibinfo {pages}
  {736–749} (\bibinfo {year} {2021})}\BibitemShut {NoStop}%
\bibitem [{\citenamefont {Hohenberg}\ and\ \citenamefont
  {Kohn}(1964)}]{Hohenberg-Kohn1964PhysRev}%
  \BibitemOpen
  \bibfield  {author} {\bibinfo {author} {\bibfnamefont {P.}~\bibnamefont
  {Hohenberg}}\ and\ \bibinfo {author} {\bibfnamefont {W.}~\bibnamefont
  {Kohn}},\ }\href@noop {} {\bibfield  {journal} {\bibinfo  {journal} {Phys.
  Rev.}\ }\textbf {\bibinfo {volume} {136}},\ \bibinfo {pages} {B864} (\bibinfo
  {year} {1964})}\BibitemShut {NoStop}%
\bibitem [{\citenamefont {Martin}\ \emph {et~al.}(2016)\citenamefont {Martin},
  \citenamefont {Reining},\ and\ \citenamefont
  {Ceperley}}]{Martin-Reining-Ceperley2016book}%
  \BibitemOpen
  \bibfield  {author} {\bibinfo {author} {\bibfnamefont {R.~M.}\ \bibnamefont
  {Martin}}, \bibinfo {author} {\bibfnamefont {L.}~\bibnamefont {Reining}}, \
  and\ \bibinfo {author} {\bibfnamefont {D.}~\bibnamefont {Ceperley}},\ }\href
  {https://doi.org/10.1017/CBO9781139050807} {\emph {\bibinfo {title}
  {Interacting Electrons Theory and Computational Approaches}}}\ (\bibinfo
  {publisher} {Cambridge University Press},\ \bibinfo {year}
  {2016})\BibitemShut {NoStop}%
\bibitem [{\citenamefont {Runge}\ and\ \citenamefont
  {Gross}(1984)}]{Runge-Gross1984PhysRevLett}%
  \BibitemOpen
  \bibfield  {author} {\bibinfo {author} {\bibfnamefont {E.}~\bibnamefont
  {Runge}}\ and\ \bibinfo {author} {\bibfnamefont {E.~K.~U.}\ \bibnamefont
  {Gross}},\ }\href {\doibase 10.1103/PhysRevLett.52.997} {\bibfield  {journal}
  {\bibinfo  {journal} {Phys. Rev. Lett.}\ }\textbf {\bibinfo {volume} {52}},\
  \bibinfo {pages} {997} (\bibinfo {year} {1984})}\BibitemShut {NoStop}%
\bibitem [{\citenamefont {Petersilka}\ \emph {et~al.}(1996)\citenamefont
  {Petersilka}, \citenamefont {Gossmann},\ and\ \citenamefont
  {Gross}}]{Petersilka1996PhysRevLett}%
  \BibitemOpen
  \bibfield  {author} {\bibinfo {author} {\bibfnamefont {M.}~\bibnamefont
  {Petersilka}}, \bibinfo {author} {\bibfnamefont {U.~J.}\ \bibnamefont
  {Gossmann}}, \ and\ \bibinfo {author} {\bibfnamefont {E.~K.~U.}\ \bibnamefont
  {Gross}},\ }\href {\doibase 10.1103/PhysRevLett.76.1212} {\bibfield
  {journal} {\bibinfo  {journal} {Phys. Rev. Lett.}\ }\textbf {\bibinfo
  {volume} {76}},\ \bibinfo {pages} {1212} (\bibinfo {year}
  {1996})}\BibitemShut {NoStop}%
\bibitem [{\citenamefont {Ullrich}(2012)}]{Ullrich2012book}%
  \BibitemOpen
  \bibfield  {author} {\bibinfo {author} {\bibfnamefont {C.~A.}\ \bibnamefont
  {Ullrich}},\ }\href@noop {} {\emph {\bibinfo {title} {Time-Dependent
  Density-Functional Theory: Concepts and Applications}}}\ (\bibinfo
  {publisher} {Oxford University Press},\ \bibinfo {year} {2012})\BibitemShut
  {NoStop}%
\bibitem [{\citenamefont {Gross}\ \emph {et~al.}(1988)\citenamefont {Gross},
  \citenamefont {Oliveira},\ and\ \citenamefont
  {Kohn}}]{Gross-Oliveira-Kohn1988PRA}%
  \BibitemOpen
  \bibfield  {author} {\bibinfo {author} {\bibfnamefont {E.~K.~U.}\
  \bibnamefont {Gross}}, \bibinfo {author} {\bibfnamefont {L.~N.}\ \bibnamefont
  {Oliveira}}, \ and\ \bibinfo {author} {\bibfnamefont {W.}~\bibnamefont
  {Kohn}},\ }\href {\doibase 10.1103/PhysRevA.37.2809} {\bibfield  {journal}
  {\bibinfo  {journal} {Phys. Rev. A}\ }\textbf {\bibinfo {volume} {37}},\
  \bibinfo {pages} {2809} (\bibinfo {year} {1988})}\BibitemShut {NoStop}%
\bibitem [{\citenamefont {Gould}\ and\ \citenamefont
  {Pittalis}(2019)}]{Gould2019PhysRevLett}%
  \BibitemOpen
  \bibfield  {author} {\bibinfo {author} {\bibfnamefont {T.}~\bibnamefont
  {Gould}}\ and\ \bibinfo {author} {\bibfnamefont {S.}~\bibnamefont
  {Pittalis}},\ }\href {\doibase 10.1103/PhysRevLett.123.016401} {\bibfield
  {journal} {\bibinfo  {journal} {Phys. Rev. Lett.}\ }\textbf {\bibinfo
  {volume} {123}},\ \bibinfo {pages} {016401} (\bibinfo {year}
  {2019})}\BibitemShut {NoStop}%
\bibitem [{\citenamefont {Cernatic}\ \emph {et~al.}(2021)\citenamefont
  {Cernatic}, \citenamefont {Senjean}, \citenamefont {Robert},\ and\
  \citenamefont {Fromager}}]{Cernatic2021TopCurrChem}%
  \BibitemOpen
  \bibfield  {author} {\bibinfo {author} {\bibfnamefont {F.}~\bibnamefont
  {Cernatic}}, \bibinfo {author} {\bibfnamefont {B.}~\bibnamefont {Senjean}},
  \bibinfo {author} {\bibfnamefont {V.}~\bibnamefont {Robert}}, \ and\ \bibinfo
  {author} {\bibfnamefont {E.}~\bibnamefont {Fromager}},\ }\href {\doibase
  10.1007/s41061-021-00359-1} {\bibfield  {journal} {\bibinfo  {journal} {Top.
  Curr. Chem. (Z)}\ }\textbf {\bibinfo {volume} {380}},\ \bibinfo {pages} {4}
  (\bibinfo {year} {2021})}\BibitemShut {NoStop}%
\bibitem [{\citenamefont {Stefanucci}\ and\ \citenamefont {van
  Leeuwen}(2013)}]{Stefanucci-vanLeeuwen2013book}%
  \BibitemOpen
  \bibfield  {author} {\bibinfo {author} {\bibfnamefont {G.}~\bibnamefont
  {Stefanucci}}\ and\ \bibinfo {author} {\bibfnamefont {R.}~\bibnamefont {van
  Leeuwen}},\ }\href {https://doi.org/10.1017/CBO9781139023979} {\emph
  {\bibinfo {title} {Nonequilibrium Many-Body Theory of Quantum Systems: A
  Modern Introduction}}}\ (\bibinfo  {publisher} {Cambridge University Press},\
  \bibinfo {year} {2013})\BibitemShut {NoStop}%
\bibitem [{\citenamefont {Hedin}(1965)}]{Hedin1965PR}%
  \BibitemOpen
  \bibfield  {author} {\bibinfo {author} {\bibfnamefont {L.}~\bibnamefont
  {Hedin}},\ }\href@noop {} {\bibfield  {journal} {\bibinfo  {journal} {Phys.
  Rev.}\ }\textbf {\bibinfo {volume} {139}},\ \bibinfo {pages} {A796} (\bibinfo
  {year} {1965})}\BibitemShut {NoStop}%
\bibitem [{\citenamefont {Reining}(2018)}]{Reining2018wcms}%
  \BibitemOpen
  \bibfield  {author} {\bibinfo {author} {\bibfnamefont {L.}~\bibnamefont
  {Reining}},\ }\href {\doibase 10.1002/wcms.1344} {\bibfield  {journal}
  {\bibinfo  {journal} {WIREs Computational Molecular Science}\ }\textbf
  {\bibinfo {volume} {8}},\ \bibinfo {pages} {e1344} (\bibinfo {year}
  {2018})}\BibitemShut {NoStop}%
\bibitem [{\citenamefont {Golze}\ \emph {et~al.}(2019)\citenamefont {Golze},
  \citenamefont {Dvorak},\ and\ \citenamefont {Rinke}}]{Golze2019FIC}%
  \BibitemOpen
  \bibfield  {author} {\bibinfo {author} {\bibfnamefont {D.}~\bibnamefont
  {Golze}}, \bibinfo {author} {\bibfnamefont {M.}~\bibnamefont {Dvorak}}, \
  and\ \bibinfo {author} {\bibfnamefont {P.}~\bibnamefont {Rinke}},\ }\href
  {\doibase 10.3389/fchem.2019.00377} {\bibfield  {journal} {\bibinfo
  {journal} {Frontiers in Chemistry}\ }\textbf {\bibinfo {volume} {7}},\
  \bibinfo {pages} {377} (\bibinfo {year} {2019})}\BibitemShut {NoStop}%
\bibitem [{\citenamefont {Onida}\ \emph {et~al.}(2002)\citenamefont {Onida},
  \citenamefont {Reining},\ and\ \citenamefont {Rubio}}]{Onida2002RMP}%
  \BibitemOpen
  \bibfield  {author} {\bibinfo {author} {\bibfnamefont {G.}~\bibnamefont
  {Onida}}, \bibinfo {author} {\bibfnamefont {L.}~\bibnamefont {Reining}}, \
  and\ \bibinfo {author} {\bibfnamefont {A.}~\bibnamefont {Rubio}},\
  }\href@noop {} {\bibfield  {journal} {\bibinfo  {journal} {Rev. Mod. Phys.}\
  }\textbf {\bibinfo {volume} {74}},\ \bibinfo {pages} {601} (\bibinfo {year}
  {2002})}\BibitemShut {NoStop}%
\bibitem [{\citenamefont {Fetter}\ and\ \citenamefont
  {Walecka}(1971)}]{fett-wale71book}%
  \BibitemOpen
  \bibfield  {author} {\bibinfo {author} {\bibfnamefont {A.~L.}\ \bibnamefont
  {Fetter}}\ and\ \bibinfo {author} {\bibfnamefont {J.~D.}\ \bibnamefont
  {Walecka}},\ }\href
  {https://books.google.it/books/about/Quantum_Theory_of_Many_particle_Systems.html?id=0wekf1s83b0C&redir_esc=y}
  {\emph {\bibinfo {title} {Quantum theory of many-particle systems}}}\
  (\bibinfo  {publisher} {McGraw-Hill, New York},\ \bibinfo {year}
  {1971})\BibitemShut {NoStop}%
\bibitem [{\citenamefont {Luttinger}\ and\ \citenamefont
  {Ward}(1960)}]{Luttinger-Ward1960PR}%
  \BibitemOpen
  \bibfield  {author} {\bibinfo {author} {\bibfnamefont {J.~M.}\ \bibnamefont
  {Luttinger}}\ and\ \bibinfo {author} {\bibfnamefont {J.~C.}\ \bibnamefont
  {Ward}},\ }\href@noop {} {\bibfield  {journal} {\bibinfo  {journal} {Phys.
  Rev.}\ }\textbf {\bibinfo {volume} {118}},\ \bibinfo {pages} {1417} (\bibinfo
  {year} {1960})}\BibitemShut {NoStop}%
\bibitem [{\citenamefont {Klein}(1961)}]{Klein1961PR}%
  \BibitemOpen
  \bibfield  {author} {\bibinfo {author} {\bibfnamefont {A.}~\bibnamefont
  {Klein}},\ }\href@noop {} {\bibfield  {journal} {\bibinfo  {journal} {Phys.
  Rev.}\ }\textbf {\bibinfo {volume} {121}},\ \bibinfo {pages} {950} (\bibinfo
  {year} {1961})}\BibitemShut {NoStop}%
\bibitem [{\citenamefont {Baym}\ and\ \citenamefont
  {Kadanoff}(1961)}]{Baym1961PhysRev}%
  \BibitemOpen
  \bibfield  {author} {\bibinfo {author} {\bibfnamefont {G.}~\bibnamefont
  {Baym}}\ and\ \bibinfo {author} {\bibfnamefont {L.~P.}\ \bibnamefont
  {Kadanoff}},\ }\href {\doibase 10.1103/PhysRev.124.287} {\bibfield  {journal}
  {\bibinfo  {journal} {Phys. Rev.}\ }\textbf {\bibinfo {volume} {124}},\
  \bibinfo {pages} {287} (\bibinfo {year} {1961})}\BibitemShut {NoStop}%
\bibitem [{\citenamefont {Almbladh}\ \emph {et~al.}(1999)\citenamefont
  {Almbladh}, \citenamefont {{von Barth}},\ and\ \citenamefont {{van
  Leeuwen}}}]{Almbladh1999InternationalJournalofModernPhysicsB}%
  \BibitemOpen
  \bibfield  {author} {\bibinfo {author} {\bibfnamefont {C.-O.}\ \bibnamefont
  {Almbladh}}, \bibinfo {author} {\bibfnamefont {U.}~\bibnamefont {{von
  Barth}}}, \ and\ \bibinfo {author} {\bibfnamefont {R.}~\bibnamefont {{van
  Leeuwen}}},\ }\href {\doibase 10.1142/s0217979299000436} {\bibfield
  {journal} {\bibinfo  {journal} {Intl. J. Mod. Phys. B}\ }\textbf {\bibinfo
  {volume} {13}},\ \bibinfo {pages} {535} (\bibinfo {year} {1999})}\BibitemShut
  {NoStop}%
\bibitem [{\citenamefont {Dahlen}\ \emph {et~al.}(2004)\citenamefont {Dahlen},
  \citenamefont {van Leeuwen},\ and\ \citenamefont {von
  Barth}}]{Dahlen2004IntJQuantumChem}%
  \BibitemOpen
  \bibfield  {author} {\bibinfo {author} {\bibfnamefont {N.~E.}\ \bibnamefont
  {Dahlen}}, \bibinfo {author} {\bibfnamefont {R.}~\bibnamefont {van Leeuwen}},
  \ and\ \bibinfo {author} {\bibfnamefont {U.}~\bibnamefont {von Barth}},\
  }\href {\doibase 10.1002/qua.20306} {\bibfield  {journal} {\bibinfo
  {journal} {Int. J. Quantum Chem.}\ }\textbf {\bibinfo {volume} {101}},\
  \bibinfo {pages} {512–519} (\bibinfo {year} {2004})}\BibitemShut {NoStop}%
\bibitem [{\citenamefont {Dahlen}\ and\ \citenamefont {von
  Barth}(2004{\natexlab{a}})}]{Dahlen2004JChemPhys}%
  \BibitemOpen
  \bibfield  {author} {\bibinfo {author} {\bibfnamefont {N.~E.}\ \bibnamefont
  {Dahlen}}\ and\ \bibinfo {author} {\bibfnamefont {U.}~\bibnamefont {von
  Barth}},\ }\href {\doibase 10.1063/1.1650307} {\bibfield  {journal} {\bibinfo
   {journal} {J. Chem. Phys.}\ }\textbf {\bibinfo {volume} {120}},\ \bibinfo
  {pages} {6826–6831} (\bibinfo {year} {2004}{\natexlab{a}})}\BibitemShut
  {NoStop}%
\bibitem [{\citenamefont {Dahlen}\ and\ \citenamefont {von
  Barth}(2004{\natexlab{b}})}]{Dahlen2004PhysRevB}%
  \BibitemOpen
  \bibfield  {author} {\bibinfo {author} {\bibfnamefont {N.~E.}\ \bibnamefont
  {Dahlen}}\ and\ \bibinfo {author} {\bibfnamefont {U.}~\bibnamefont {von
  Barth}},\ }\href {\doibase 10.1103/physrevb.69.195102} {\bibfield  {journal}
  {\bibinfo  {journal} {Phys. Rev. B}\ }\textbf {\bibinfo {volume} {69}},\
  \bibinfo {pages} {195102} (\bibinfo {year} {2004}{\natexlab{b}})}\BibitemShut
  {NoStop}%
\bibitem [{\citenamefont {Dahlen}\ \emph {et~al.}(2006)\citenamefont {Dahlen},
  \citenamefont {van Leeuwen},\ and\ \citenamefont {von
  Barth}}]{Dahlen2006PhysRevA}%
  \BibitemOpen
  \bibfield  {author} {\bibinfo {author} {\bibfnamefont {N.~E.}\ \bibnamefont
  {Dahlen}}, \bibinfo {author} {\bibfnamefont {R.}~\bibnamefont {van Leeuwen}},
  \ and\ \bibinfo {author} {\bibfnamefont {U.}~\bibnamefont {von Barth}},\
  }\href {\doibase 10.1103/physreva.73.012511} {\bibfield  {journal} {\bibinfo
  {journal} {Phys. Rev. A}\ }\textbf {\bibinfo {volume} {73}},\ \bibinfo
  {pages} {012511} (\bibinfo {year} {2006})}\BibitemShut {NoStop}%
\bibitem [{\citenamefont {Puig~von Friesen}\ \emph {et~al.}(2010)\citenamefont
  {Puig~von Friesen}, \citenamefont {Verdozzi},\ and\ \citenamefont
  {Almbladh}}]{Friesen2010PhysRevB}%
  \BibitemOpen
  \bibfield  {author} {\bibinfo {author} {\bibfnamefont {M.}~\bibnamefont
  {Puig~von Friesen}}, \bibinfo {author} {\bibfnamefont {C.}~\bibnamefont
  {Verdozzi}}, \ and\ \bibinfo {author} {\bibfnamefont {C.-O.}\ \bibnamefont
  {Almbladh}},\ }\href {\doibase 10.1103/physrevb.82.155108} {\bibfield
  {journal} {\bibinfo  {journal} {Phys. Rev. B}\ }\textbf {\bibinfo {volume}
  {82}},\ \bibinfo {pages} {155108} (\bibinfo {year} {2010})}\BibitemShut
  {NoStop}%
\bibitem [{\citenamefont {Di~Sabatino}\ \emph {et~al.}(2021)\citenamefont
  {Di~Sabatino}, \citenamefont {Loos},\ and\ \citenamefont
  {Romaniello}}]{Sabatino2021FrontChem}%
  \BibitemOpen
  \bibfield  {author} {\bibinfo {author} {\bibfnamefont {S.}~\bibnamefont
  {Di~Sabatino}}, \bibinfo {author} {\bibfnamefont {P.-F.}\ \bibnamefont
  {Loos}}, \ and\ \bibinfo {author} {\bibfnamefont {P.}~\bibnamefont
  {Romaniello}},\ }\href {\doibase 10.3389/fchem.2021.751054} {\bibfield
  {journal} {\bibinfo  {journal} {Front. Chem.}\ }\textbf {\bibinfo {volume}
  {9}},\ \bibinfo {pages} {751054} (\bibinfo {year} {2021})}\BibitemShut
  {NoStop}%
\bibitem [{\citenamefont {Holm}\ and\ \citenamefont
  {Aryasetiawan}(2000)}]{Holm2000PhysRevB}%
  \BibitemOpen
  \bibfield  {author} {\bibinfo {author} {\bibfnamefont {B.}~\bibnamefont
  {Holm}}\ and\ \bibinfo {author} {\bibfnamefont {F.}~\bibnamefont
  {Aryasetiawan}},\ }\href {\doibase 10.1103/physrevb.62.4858} {\bibfield
  {journal} {\bibinfo  {journal} {Phys. Rev. B}\ }\textbf {\bibinfo {volume}
  {62}},\ \bibinfo {pages} {4858-4865} (\bibinfo {year} {2000})}\BibitemShut
  {NoStop}%
\bibitem [{\citenamefont {García-González}\ and\ \citenamefont
  {Godby}(2001)}]{Garcia-Gonzalez2001PhysRevB}%
  \BibitemOpen
  \bibfield  {author} {\bibinfo {author} {\bibfnamefont {P.}~\bibnamefont
  {Garc{\'\i}a-Gonz{\'a}lez}}\ and\ \bibinfo {author} {\bibfnamefont {R.~W.}\
  \bibnamefont {Godby}},\ }\href {\doibase 10.1103/physrevb.63.075112}
  {\bibfield  {journal} {\bibinfo  {journal} {Phys. Rev. B}\ }\textbf {\bibinfo
  {volume} {63}},\ \bibinfo {pages} {075112} (\bibinfo {year}
  {2001})}\BibitemShut {NoStop}%
\bibitem [{\citenamefont {Chiarotti}\ \emph {et~al.}(2022)\citenamefont
  {Chiarotti}, \citenamefont {Marzari},\ and\ \citenamefont
  {Ferretti}}]{Chiarotti2022PRR}%
  \BibitemOpen
  \bibfield  {author} {\bibinfo {author} {\bibfnamefont {T.}~\bibnamefont
  {Chiarotti}}, \bibinfo {author} {\bibfnamefont {N.}~\bibnamefont {Marzari}},
  \ and\ \bibinfo {author} {\bibfnamefont {A.}~\bibnamefont {Ferretti}},\
  }\href {\doibase 10.1103/physrevresearch.4.013242} {\bibfield  {journal}
  {\bibinfo  {journal} {Phys. Rev. Res.}\ }\textbf {\bibinfo {volume} {4}},\
  \bibinfo {pages} {013242} (\bibinfo {year} {2022})}\BibitemShut {NoStop}%
\bibitem [{\citenamefont {Casida}(1995)}]{Casida1995PhysRevA}%
  \BibitemOpen
  \bibfield  {author} {\bibinfo {author} {\bibfnamefont {M.~E.}\ \bibnamefont
  {Casida}},\ }\href {\doibase 10.1103/physreva.51.2005} {\bibfield  {journal}
  {\bibinfo  {journal} {Phys. Rev. A}\ }\textbf {\bibinfo {volume} {51}},\
  \bibinfo {pages} {2005-2013} (\bibinfo {year} {1995})}\BibitemShut
  {NoStop}%
\bibitem [{\citenamefont {K{\"u}mmel}\ and\ \citenamefont
  {Kronik}(2008)}]{Kummel-Kronik2008RMP}%
  \BibitemOpen
  \bibfield  {author} {\bibinfo {author} {\bibfnamefont {S.}~\bibnamefont
  {K{\"u}mmel}}\ and\ \bibinfo {author} {\bibfnamefont {L.}~\bibnamefont
  {Kronik}},\ }\href {\doibase 10.1103/RevModPhys.80.3} {\bibfield  {journal}
  {\bibinfo  {journal} {Rev. Mod. Phys.}\ }\textbf {\bibinfo {volume} {80}},\
  \bibinfo {pages} {3} (\bibinfo {year} {2008})}\BibitemShut {NoStop}%
\bibitem [{\citenamefont {Sham}\ and\ \citenamefont
  {Schl\"uter}(1983)}]{Sham-Schluter1983PRL}%
  \BibitemOpen
  \bibfield  {author} {\bibinfo {author} {\bibfnamefont {L.~J.}\ \bibnamefont
  {Sham}}\ and\ \bibinfo {author} {\bibfnamefont {M.}~\bibnamefont
  {Schl\"uter}},\ }\href {\doibase 10.1103/PhysRevLett.51.1888} {\bibfield
  {journal} {\bibinfo  {journal} {Phys. Rev. Lett.}\ }\textbf {\bibinfo
  {volume} {51}},\ \bibinfo {pages} {1888} (\bibinfo {year}
  {1983})}\BibitemShut {NoStop}%
\bibitem [{\citenamefont {Godby}\ \emph {et~al.}(1987)\citenamefont {Godby},
  \citenamefont {Schl\"uter},\ and\ \citenamefont
  {Sham}}]{Godby-Schluter-Sham1987PRB}%
  \BibitemOpen
  \bibfield  {author} {\bibinfo {author} {\bibfnamefont {R.~W.}\ \bibnamefont
  {Godby}}, \bibinfo {author} {\bibfnamefont {M.}~\bibnamefont {Schl\"uter}}, \
  and\ \bibinfo {author} {\bibfnamefont {L.~J.}\ \bibnamefont {Sham}},\ }\href
  {\doibase 10.1103/PhysRevB.36.6497} {\bibfield  {journal} {\bibinfo
  {journal} {Phys. Rev. B}\ }\textbf {\bibinfo {volume} {36}},\ \bibinfo
  {pages} {6497} (\bibinfo {year} {1987})}\BibitemShut {NoStop}%
\bibitem [{\citenamefont {Ismail-Beigi}(2010)}]{Ismail-Beigi2010PhysRevB}%
  \BibitemOpen
  \bibfield  {author} {\bibinfo {author} {\bibfnamefont {S.}~\bibnamefont
  {Ismail-Beigi}},\ }\href {\doibase 10.1103/physrevb.81.195126} {\bibfield
  {journal} {\bibinfo  {journal} {Phys. Rev. B}\ }\textbf {\bibinfo {volume}
  {81}},\ \bibinfo {pages} {195126} (\bibinfo {year} {2010})}\BibitemShut
  {NoStop}%
\bibitem [{\citenamefont {Ren}\ \emph {et~al.}(2012)\citenamefont {Ren},
  \citenamefont {Rinke}, \citenamefont {Joas},\ and\ \citenamefont
  {Scheffler}}]{Ren2012JMaterSci}%
  \BibitemOpen
  \bibfield  {author} {\bibinfo {author} {\bibfnamefont {X.}~\bibnamefont
  {Ren}}, \bibinfo {author} {\bibfnamefont {P.}~\bibnamefont {Rinke}}, \bibinfo
  {author} {\bibfnamefont {C.}~\bibnamefont {Joas}}, \ and\ \bibinfo {author}
  {\bibfnamefont {M.}~\bibnamefont {Scheffler}},\ }\href {\doibase
  10.1007/s10853-012-6570-4} {\bibfield  {journal} {\bibinfo  {journal} {J.
  Mater. Sci.}\ }\textbf {\bibinfo {volume} {47}},\ \bibinfo {pages}
  {7447–7471} (\bibinfo {year} {2012})}\BibitemShut {NoStop}%
\bibitem [{\citenamefont {Paier}\ \emph {et~al.}(2012)\citenamefont {Paier},
  \citenamefont {Ren}, \citenamefont {Rinke}, \citenamefont {Scuseria},
  \citenamefont {Grüneis}, \citenamefont {Kresse},\ and\ \citenamefont
  {Scheffler}}]{Paier2012NewJPhys}%
  \BibitemOpen
  \bibfield  {author} {\bibinfo {author} {\bibfnamefont {J.}~\bibnamefont
  {Paier}}, \bibinfo {author} {\bibfnamefont {X.}~\bibnamefont {Ren}}, \bibinfo
  {author} {\bibfnamefont {P.}~\bibnamefont {Rinke}}, \bibinfo {author}
  {\bibfnamefont {G.~E.}\ \bibnamefont {Scuseria}}, \bibinfo {author}
  {\bibfnamefont {A.}~\bibnamefont {Grüneis}}, \bibinfo {author}
  {\bibfnamefont {G.}~\bibnamefont {Kresse}}, \ and\ \bibinfo {author}
  {\bibfnamefont {M.}~\bibnamefont {Scheffler}},\ }\href {\doibase
  10.1088/1367-2630/14/4/043002} {\bibfield  {journal} {\bibinfo  {journal}
  {New J. Phys.}\ }\textbf {\bibinfo {volume} {14}},\ \bibinfo {pages} {043002}
  (\bibinfo {year} {2012})}\BibitemShut {NoStop}%
\bibitem [{\citenamefont {Hellgren}\ \emph {et~al.}(2018)\citenamefont
  {Hellgren}, \citenamefont {Colonna},\ and\ \citenamefont
  {de~Gironcoli}}]{Hellgren2018PhysRevB}%
  \BibitemOpen
  \bibfield  {author} {\bibinfo {author} {\bibfnamefont {M.}~\bibnamefont
  {Hellgren}}, \bibinfo {author} {\bibfnamefont {N.}~\bibnamefont {Colonna}}, \
  and\ \bibinfo {author} {\bibfnamefont {S.}~\bibnamefont {de~Gironcoli}},\
  }\href {\doibase 10.1103/physrevb.98.045117} {\bibfield  {journal} {\bibinfo
  {journal} {Phys. Rev. B}\ }\textbf {\bibinfo {volume} {98}},\ \bibinfo
  {pages} {045117} (\bibinfo {year} {2018})}\BibitemShut {NoStop}%
\bibitem [{\citenamefont {Kotliar}\ \emph {et~al.}(2006)\citenamefont
  {Kotliar}, \citenamefont {Savrasov}, \citenamefont {Haule}, \citenamefont
  {Oudovenko}, \citenamefont {Parcollet},\ and\ \citenamefont
  {Marianetti}}]{Kotliar2006RevModPhys}%
  \BibitemOpen
  \bibfield  {author} {\bibinfo {author} {\bibfnamefont {G.}~\bibnamefont
  {Kotliar}}, \bibinfo {author} {\bibfnamefont {S.~Y.}\ \bibnamefont
  {Savrasov}}, \bibinfo {author} {\bibfnamefont {K.}~\bibnamefont {Haule}},
  \bibinfo {author} {\bibfnamefont {V.~S.}\ \bibnamefont {Oudovenko}}, \bibinfo
  {author} {\bibfnamefont {O.}~\bibnamefont {Parcollet}}, \ and\ \bibinfo
  {author} {\bibfnamefont {C.~A.}\ \bibnamefont {Marianetti}},\ }\href
  {\doibase 10.1103/revmodphys.78.865} {\bibfield  {journal} {\bibinfo
  {journal} {Rev. Mod. Phys.}\ }\textbf {\bibinfo {volume} {78}},\ \bibinfo
  {pages} {865-951} (\bibinfo {year} {2006})}\BibitemShut {NoStop}%
\bibitem [{\citenamefont {Gatti}\ \emph {et~al.}(2007)\citenamefont {Gatti},
  \citenamefont {Bruneval}, \citenamefont {Olevano},\ and\ \citenamefont
  {Reining}}]{Gatti2007PhysRevLett}%
  \BibitemOpen
  \bibfield  {author} {\bibinfo {author} {\bibfnamefont {M.}~\bibnamefont
  {Gatti}}, \bibinfo {author} {\bibfnamefont {F.}~\bibnamefont {Bruneval}},
  \bibinfo {author} {\bibfnamefont {V.}~\bibnamefont {Olevano}}, \ and\
  \bibinfo {author} {\bibfnamefont {L.}~\bibnamefont {Reining}},\ }\href
  {\doibase 10.1103/physrevlett.99.266402} {\bibfield  {journal} {\bibinfo
  {journal} {Phys. Rev. Lett.}\ }\textbf {\bibinfo {volume} {99}},\ \bibinfo
  {pages} {266402} (\bibinfo {year} {2007})}\BibitemShut {NoStop}%
\bibitem [{\citenamefont {Ferretti}\ \emph {et~al.}(2014)\citenamefont
  {Ferretti}, \citenamefont {Dabo}, \citenamefont {Cococcioni},\ and\
  \citenamefont {Marzari}}]{Ferretti2014PhysRevB}%
  \BibitemOpen
  \bibfield  {author} {\bibinfo {author} {\bibfnamefont {A.}~\bibnamefont
  {Ferretti}}, \bibinfo {author} {\bibfnamefont {I.}~\bibnamefont {Dabo}},
  \bibinfo {author} {\bibfnamefont {M.}~\bibnamefont {Cococcioni}}, \ and\
  \bibinfo {author} {\bibfnamefont {N.}~\bibnamefont {Marzari}},\ }\href
  {\doibase 10.1103/physrevb.89.195134} {\bibfield  {journal} {\bibinfo
  {journal} {Phys. Rev. B}\ }\textbf {\bibinfo {volume} {89}},\ \bibinfo
  {pages} {195134} (\bibinfo {year} {2014})}\BibitemShut {NoStop}%
\bibitem [{\citenamefont {Dabo}\ \emph {et~al.}(2009)\citenamefont {Dabo},
  \citenamefont {Cococcioni},\ and\ \citenamefont {Marzari}}]{Dabo2009arXiv}%
  \BibitemOpen
  \bibfield  {author} {\bibinfo {author} {\bibfnamefont {I.}~\bibnamefont
  {Dabo}}, \bibinfo {author} {\bibfnamefont {M.}~\bibnamefont {Cococcioni}}, \
  and\ \bibinfo {author} {\bibfnamefont {N.}~\bibnamefont {Marzari}},\ }\href
  {http://arxiv.org/abs/0901.2637v1} {\bibfield  {journal} {\bibinfo  {journal}
  {arXiv}\ } (\bibinfo {year} {2009})},\ \Eprint
  {http://arxiv.org/abs/0901.2637v1} {arXiv:0901.2637v1 [cond-mat.mtrl-sci]}
  \BibitemShut {NoStop}%
\bibitem [{\citenamefont {Dabo}\ \emph {et~al.}(2010)\citenamefont {Dabo},
  \citenamefont {Ferretti}, \citenamefont {Poilvert}, \citenamefont {Li},
  \citenamefont {Marzari},\ and\ \citenamefont
  {Cococcioni}}]{Dabo2010PhysRevB}%
  \BibitemOpen
  \bibfield  {author} {\bibinfo {author} {\bibfnamefont {I.}~\bibnamefont
  {Dabo}}, \bibinfo {author} {\bibfnamefont {A.}~\bibnamefont {Ferretti}},
  \bibinfo {author} {\bibfnamefont {N.}~\bibnamefont {Poilvert}}, \bibinfo
  {author} {\bibfnamefont {Y.}~\bibnamefont {Li}}, \bibinfo {author}
  {\bibfnamefont {N.}~\bibnamefont {Marzari}}, \ and\ \bibinfo {author}
  {\bibfnamefont {M.}~\bibnamefont {Cococcioni}},\ }\href {\doibase
  10.1103/physrevb.82.115121} {\bibfield  {journal} {\bibinfo  {journal} {Phys.
  Rev. B}\ }\textbf {\bibinfo {volume} {82}},\ \bibinfo {pages} {115121}
  (\bibinfo {year} {2010})}\BibitemShut {NoStop}%
\bibitem [{\citenamefont {Nguyen}\ \emph {et~al.}(2018)\citenamefont {Nguyen},
  \citenamefont {Colonna}, \citenamefont {Ferretti},\ and\ \citenamefont
  {Marzari}}]{Nguyen2018PhysRevX}%
  \BibitemOpen
  \bibfield  {author} {\bibinfo {author} {\bibfnamefont {N.~L.}\ \bibnamefont
  {Nguyen}}, \bibinfo {author} {\bibfnamefont {N.}~\bibnamefont {Colonna}},
  \bibinfo {author} {\bibfnamefont {A.}~\bibnamefont {Ferretti}}, \ and\
  \bibinfo {author} {\bibfnamefont {N.}~\bibnamefont {Marzari}},\ }\href
  {\doibase 10.1103/physrevx.8.021051} {\bibfield  {journal} {\bibinfo
  {journal} {Phys. Rev. X}\ }\textbf {\bibinfo {volume} {8}},\ \bibinfo {pages}
  {021051} (\bibinfo {year} {2018})}\BibitemShut {NoStop}%
\bibitem [{\citenamefont {Meir}\ and\ \citenamefont
  {Wingreen}(1992)}]{Meir1992PhysRevLett}%
  \BibitemOpen
  \bibfield  {author} {\bibinfo {author} {\bibfnamefont {Y.}~\bibnamefont
  {Meir}}\ and\ \bibinfo {author} {\bibfnamefont {N.~S.}\ \bibnamefont
  {Wingreen}},\ }\href {\doibase 10.1103/physrevlett.68.2512} {\bibfield
  {journal} {\bibinfo  {journal} {Phys. Rev. Lett.}\ }\textbf {\bibinfo
  {volume} {68}},\ \bibinfo {pages} {2512-2515} (\bibinfo {year}
  {1992})}\BibitemShut {NoStop}%
\bibitem [{\citenamefont {Haug}\ and\ \citenamefont
  {Jauho}(1996)}]{Haug-Jauho1996book}%
  \BibitemOpen
  \bibfield  {author} {\bibinfo {author} {\bibfnamefont {H.}~\bibnamefont
  {Haug}}\ and\ \bibinfo {author} {\bibfnamefont {A.-P.}\ \bibnamefont
  {Jauho}},\ }\href {https://doi.org/10.1007/978-3-540-73564-9} {\emph
  {\bibinfo {title} {Transport and Optics of Semiconductors}}}\ (\bibinfo
  {publisher} {Springer, Berlin},\ \bibinfo {year} {1996})\BibitemShut
  {NoStop}%
\bibitem [{\citenamefont {{Buongiorno Nardelli}}(1999)}]{Nardelli1999PhysRevB}%
  \BibitemOpen
  \bibfield  {author} {\bibinfo {author} {\bibfnamefont {M.}~\bibnamefont
  {{Buongiorno Nardelli}}},\ }\href {\doibase 10.1103/physrevb.60.7828}
  {\bibfield  {journal} {\bibinfo  {journal} {Phys. Rev. B}\ }\textbf {\bibinfo
  {volume} {60}},\ \bibinfo {pages} {7828-7833} (\bibinfo {year}
  {1999})}\BibitemShut {NoStop}%
\bibitem [{\citenamefont {Brandbyge}\ \emph {et~al.}(2002)\citenamefont
  {Brandbyge}, \citenamefont {Mozos}, \citenamefont {Ordej\'on}, \citenamefont
  {Taylor},\ and\ \citenamefont {Stokbro}}]{Brandbyge2002PRB}%
  \BibitemOpen
  \bibfield  {author} {\bibinfo {author} {\bibfnamefont {M.}~\bibnamefont
  {Brandbyge}}, \bibinfo {author} {\bibfnamefont {J.-L.}\ \bibnamefont
  {Mozos}}, \bibinfo {author} {\bibfnamefont {P.}~\bibnamefont {Ordej\'on}},
  \bibinfo {author} {\bibfnamefont {J.}~\bibnamefont {Taylor}}, \ and\ \bibinfo
  {author} {\bibfnamefont {K.}~\bibnamefont {Stokbro}},\ }\href {\doibase
  10.1103/PhysRevB.65.165401} {\bibfield  {journal} {\bibinfo  {journal} {Phys.
  Rev. B}\ }\textbf {\bibinfo {volume} {65}},\ \bibinfo {pages} {165401}
  (\bibinfo {year} {2002})}\BibitemShut {NoStop}%
\bibitem [{\citenamefont {Calzolari}\ \emph {et~al.}(2004)\citenamefont
  {Calzolari}, \citenamefont {Marzari}, \citenamefont {Souza},\ and\
  \citenamefont {Buongiorno~Nardelli}}]{Calzolari2004PhysRevB}%
  \BibitemOpen
  \bibfield  {author} {\bibinfo {author} {\bibfnamefont {A.}~\bibnamefont
  {Calzolari}}, \bibinfo {author} {\bibfnamefont {N.}~\bibnamefont {Marzari}},
  \bibinfo {author} {\bibfnamefont {I.}~\bibnamefont {Souza}}, \ and\ \bibinfo
  {author} {\bibfnamefont {M.}~\bibnamefont {Buongiorno~Nardelli}},\ }\href
  {\doibase 10.1103/physrevb.69.035108} {\bibfield  {journal} {\bibinfo
  {journal} {Phys. Rev. B}\ }\textbf {\bibinfo {volume} {69}},\ \bibinfo
  {pages} {035108} (\bibinfo {year} {2004})}\BibitemShut {NoStop}%
\bibitem [{\citenamefont {Chiarotti}\ \emph {et~al.}(2023)\citenamefont
  {Chiarotti}, \citenamefont {Ferretti},\ and\ \citenamefont
  {Marzari}}]{Chiarotti2023arXiv}%
  \BibitemOpen
  \bibfield  {author} {\bibinfo {author} {\bibfnamefont {T.}~\bibnamefont
  {Chiarotti}}, \bibinfo {author} {\bibfnamefont {A.}~\bibnamefont {Ferretti}},
  \ and\ \bibinfo {author} {\bibfnamefont {N.}~\bibnamefont {Marzari}},\ }\href
  {http://arxiv.org/abs/2302.12193v1} {\bibfield  {journal} {\bibinfo
  {journal} {arXiv}\ } (\bibinfo {year} {2023})},\ \Eprint
  {http://arxiv.org/abs/2302.12193v1} {arXiv:2302.12193v1 [cond-mat.mtrl-sci]}
  \BibitemShut {NoStop}%
\bibitem [{\citenamefont {Chiarotti}(2023)}]{Chiarotti2023PhD}%
  \BibitemOpen
  \bibfield  {author} {\bibinfo {author} {\bibfnamefont {T.}~\bibnamefont
  {Chiarotti}},\ }\emph {\bibinfo {title} {Spectral and thermodynamic
  properties of interacting electrons with dynamical functionals}},\ \href@noop
  {} {Ph.D. thesis},\ \bibinfo  {school} {EDMX Doctoral Program, École
  polytechnique fédérale de Lausanne (EPFL), Switzerland} (\bibinfo {year}
  {2023})\BibitemShut {NoStop}%
\bibitem [{\citenamefont {Engel}\ \emph {et~al.}(1991)\citenamefont {Engel},
  \citenamefont {Farid}, \citenamefont {Nex},\ and\ \citenamefont
  {March}}]{Engel1991PhysRevB}%
  \BibitemOpen
  \bibfield  {author} {\bibinfo {author} {\bibfnamefont {G.~E.}\ \bibnamefont
  {Engel}}, \bibinfo {author} {\bibfnamefont {B.}~\bibnamefont {Farid}},
  \bibinfo {author} {\bibfnamefont {C.~M.~M.}\ \bibnamefont {Nex}}, \ and\
  \bibinfo {author} {\bibfnamefont {N.~H.}\ \bibnamefont {March}},\ }\href
  {\doibase 10.1103/physrevb.44.13356} {\bibfield  {journal} {\bibinfo
  {journal} {Phys. Rev. B}\ }\textbf {\bibinfo {volume} {44}},\ \bibinfo
  {pages} {13356-13373} (\bibinfo {year} {1991})}\BibitemShut {NoStop}%
\bibitem [{\citenamefont {Rojas}\ \emph {et~al.}(1995)\citenamefont {Rojas},
  \citenamefont {Godby},\ and\ \citenamefont {Needs}}]{Rojas1995PhysRevLett}%
  \BibitemOpen
  \bibfield  {author} {\bibinfo {author} {\bibfnamefont {H.~N.}\ \bibnamefont
  {Rojas}}, \bibinfo {author} {\bibfnamefont {R.~W.}\ \bibnamefont {Godby}}, \
  and\ \bibinfo {author} {\bibfnamefont {R.~J.}\ \bibnamefont {Needs}},\ }\href
  {\doibase 10.1103/physrevlett.74.1827} {\bibfield  {journal} {\bibinfo
  {journal} {Phys. Rev. Lett.}\ }\textbf {\bibinfo {volume} {74}},\ \bibinfo
  {pages} {1827-1830} (\bibinfo {year} {1995})}\BibitemShut {NoStop}%
\bibitem [{\citenamefont {G\"uttel}\ and\ \citenamefont
  {Tisseur}(2017)}]{Guttel2017ActaNumerica}%
  \BibitemOpen
  \bibfield  {author} {\bibinfo {author} {\bibfnamefont {S.}~\bibnamefont
  {G\"uttel}}\ and\ \bibinfo {author} {\bibfnamefont {F.}~\bibnamefont
  {Tisseur}},\ }\href {\doibase 10.1017/s0962492917000034} {\bibfield
  {journal} {\bibinfo  {journal} {Acta Numerica}\ }\textbf {\bibinfo {volume}
  {26}},\ \bibinfo {pages} {1–94} (\bibinfo {year} {2017})}\BibitemShut
  {NoStop}%
\bibitem [{\citenamefont {Leon}\ \emph {et~al.}(2021)\citenamefont {Leon},
  \citenamefont {Cardoso}, \citenamefont {Chiarotti}, \citenamefont {Varsano},
  \citenamefont {Molinari},\ and\ \citenamefont {Ferretti}}]{Leon2021PhysRevB}%
  \BibitemOpen
  \bibfield  {author} {\bibinfo {author} {\bibfnamefont {D.~A.}\ \bibnamefont
  {Leon}}, \bibinfo {author} {\bibfnamefont {C.}~\bibnamefont {Cardoso}},
  \bibinfo {author} {\bibfnamefont {T.}~\bibnamefont {Chiarotti}}, \bibinfo
  {author} {\bibfnamefont {D.}~\bibnamefont {Varsano}}, \bibinfo {author}
  {\bibfnamefont {E.}~\bibnamefont {Molinari}}, \ and\ \bibinfo {author}
  {\bibfnamefont {A.}~\bibnamefont {Ferretti}},\ }\href {\doibase
  10.1103/physrevb.104.115157} {\bibfield  {journal} {\bibinfo  {journal}
  {Phys. Rev. B}\ }\textbf {\bibinfo {volume} {104}},\ \bibinfo {pages}
  {115157} (\bibinfo {year} {2021})}\BibitemShut {NoStop}%
\bibitem [{\citenamefont {Leon}\ \emph {et~al.}(2023)\citenamefont {Leon},
  \citenamefont {Ferretti}, \citenamefont {Varsano}, \citenamefont {Molinari},\
  and\ \citenamefont {Cardoso}}]{Leon2023PhysRevB}%
  \BibitemOpen
  \bibfield  {author} {\bibinfo {author} {\bibfnamefont {D.~A.}\ \bibnamefont
  {Leon}}, \bibinfo {author} {\bibfnamefont {A.}~\bibnamefont {Ferretti}},
  \bibinfo {author} {\bibfnamefont {D.}~\bibnamefont {Varsano}}, \bibinfo
  {author} {\bibfnamefont {E.}~\bibnamefont {Molinari}}, \ and\ \bibinfo
  {author} {\bibfnamefont {C.}~\bibnamefont {Cardoso}},\ }\href {\doibase
  10.1103/physrevb.107.155130} {\bibfield  {journal} {\bibinfo  {journal}
  {Phys. Rev. B}\ }\textbf {\bibinfo {volume} {107}},\ \bibinfo {pages}
  {155130} (\bibinfo {year} {2023})}\BibitemShut {NoStop}%
\bibitem [{\citenamefont {Stefanucci}\ \emph {et~al.}(2014)\citenamefont
  {Stefanucci}, \citenamefont {Pavlyukh}, \citenamefont {Uimonen},\ and\
  \citenamefont {van Leeuwen}}]{Stefanucci2014PhysRevB}%
  \BibitemOpen
  \bibfield  {author} {\bibinfo {author} {\bibfnamefont {G.}~\bibnamefont
  {Stefanucci}}, \bibinfo {author} {\bibfnamefont {Y.}~\bibnamefont
  {Pavlyukh}}, \bibinfo {author} {\bibfnamefont {A.-M.}\ \bibnamefont
  {Uimonen}}, \ and\ \bibinfo {author} {\bibfnamefont {R.}~\bibnamefont {van
  Leeuwen}},\ }\href {\doibase 10.1103/physrevb.90.115134} {\bibfield
  {journal} {\bibinfo  {journal} {Phys. Rev. B}\ }\textbf {\bibinfo {volume}
  {90}},\ \bibinfo {pages} {115134} (\bibinfo {year} {2014})}\BibitemShut
  {NoStop}%
\bibitem [{\citenamefont {Savrasov}\ \emph {et~al.}(2006)\citenamefont
  {Savrasov}, \citenamefont {Haule},\ and\ \citenamefont
  {Kotliar}}]{Savrasov2006PhysRevLett}%
  \BibitemOpen
  \bibfield  {author} {\bibinfo {author} {\bibfnamefont {S.~Y.}\ \bibnamefont
  {Savrasov}}, \bibinfo {author} {\bibfnamefont {K.}~\bibnamefont {Haule}}, \
  and\ \bibinfo {author} {\bibfnamefont {G.}~\bibnamefont {Kotliar}},\ }\href
  {\doibase 10.1103/physrevlett.96.036404} {\bibfield  {journal} {\bibinfo
  {journal} {Phys. Rev. Lett.}\ }\textbf {\bibinfo {volume} {96}},\ \bibinfo
  {pages} {036404} (\bibinfo {year} {2006})}\BibitemShut {NoStop}%
\bibitem [{\citenamefont {Budich}\ \emph {et~al.}(2012)\citenamefont {Budich},
  \citenamefont {Thomale}, \citenamefont {Li}, \citenamefont {Laubach},\ and\
  \citenamefont {Zhang}}]{Budich2012PhysRevB}%
  \BibitemOpen
  \bibfield  {author} {\bibinfo {author} {\bibfnamefont {J.~C.}\ \bibnamefont
  {Budich}}, \bibinfo {author} {\bibfnamefont {R.}~\bibnamefont {Thomale}},
  \bibinfo {author} {\bibfnamefont {G.}~\bibnamefont {Li}}, \bibinfo {author}
  {\bibfnamefont {M.}~\bibnamefont {Laubach}}, \ and\ \bibinfo {author}
  {\bibfnamefont {S.-C.}\ \bibnamefont {Zhang}},\ }\href {\doibase
  10.1103/physrevb.86.201407} {\bibfield  {journal} {\bibinfo  {journal} {Phys.
  Rev. B}\ }\textbf {\bibinfo {volume} {86}},\ \bibinfo {pages} {201407(R)}
  (\bibinfo {year} {2012})}\BibitemShut {NoStop}%
\bibitem [{\citenamefont {Wang}\ \emph {et~al.}(2012)\citenamefont {Wang},
  \citenamefont {Jiang}, \citenamefont {Dai},\ and\ \citenamefont
  {Xie}}]{Wang2012PhysRevB}%
  \BibitemOpen
  \bibfield  {author} {\bibinfo {author} {\bibfnamefont {L.}~\bibnamefont
  {Wang}}, \bibinfo {author} {\bibfnamefont {H.}~\bibnamefont {Jiang}},
  \bibinfo {author} {\bibfnamefont {X.}~\bibnamefont {Dai}}, \ and\ \bibinfo
  {author} {\bibfnamefont {X.~C.}\ \bibnamefont {Xie}},\ }\href {\doibase
  10.1103/physrevb.85.235135} {\bibfield  {journal} {\bibinfo  {journal} {Phys.
  Rev. B}\ }\textbf {\bibinfo {volume} {85}},\ \bibinfo {pages} {235135}
  (\bibinfo {year} {2012})}\BibitemShut {NoStop}%
\bibitem [{\citenamefont {Bintrim}\ and\ \citenamefont
  {Berkelbach}(2021)}]{Bintrim2021JChemPhys}%
  \BibitemOpen
  \bibfield  {author} {\bibinfo {author} {\bibfnamefont {S.~J.}\ \bibnamefont
  {Bintrim}}\ and\ \bibinfo {author} {\bibfnamefont {T.~C.}\ \bibnamefont
  {Berkelbach}},\ }\href {\doibase 10.1063/5.0035141} {\bibfield  {journal}
  {\bibinfo  {journal} {J. Chem. Phys.}\ }\textbf {\bibinfo {volume} {154}},\
  \bibinfo {pages} {041101} (\bibinfo {year} {2021})}\BibitemShut {NoStop}%
\bibitem [{\citenamefont {Bintrim}\ and\ \citenamefont
  {Berkelbach}(2022)}]{Bintrim2022JChemPhys}%
  \BibitemOpen
  \bibfield  {author} {\bibinfo {author} {\bibfnamefont {S.~J.}\ \bibnamefont
  {Bintrim}}\ and\ \bibinfo {author} {\bibfnamefont {T.~C.}\ \bibnamefont
  {Berkelbach}},\ }\href {\doibase 10.1063/5.0074434} {\bibfield  {journal}
  {\bibinfo  {journal} {J. Chem. Phys.}\ }\textbf {\bibinfo {volume} {156}},\
  \bibinfo {pages} {044114} (\bibinfo {year} {2022})}\BibitemShut {NoStop}%
\bibitem [{\citenamefont {Brualdi}\ and\ \citenamefont
  {Schneider}(1983)}]{Brualdi1983LinearAlgebraanditsApplications}%
  \BibitemOpen
  \bibfield  {author} {\bibinfo {author} {\bibfnamefont {R.~A.}\ \bibnamefont
  {Brualdi}}\ and\ \bibinfo {author} {\bibfnamefont {H.}~\bibnamefont
  {Schneider}},\ }\href {\doibase 10.1016/0024-3795(83)80049-4} {\bibfield
  {journal} {\bibinfo  {journal} {Linear Algebra and its Applications}\
  }\textbf {\bibinfo {volume} {52–53}},\ \bibinfo {pages} {769} (\bibinfo
  {year} {1983})}\BibitemShut {NoStop}%
\bibitem [{\citenamefont {Aryasetiawan}\ \emph {et~al.}(2002)\citenamefont
  {Aryasetiawan}, \citenamefont {Miyake},\ and\ \citenamefont
  {Terakura}}]{Aryasetiawan2002PhysRevLett}%
  \BibitemOpen
  \bibfield  {author} {\bibinfo {author} {\bibfnamefont {F.}~\bibnamefont
  {Aryasetiawan}}, \bibinfo {author} {\bibfnamefont {T.}~\bibnamefont
  {Miyake}}, \ and\ \bibinfo {author} {\bibfnamefont {K.}~\bibnamefont
  {Terakura}},\ }\href {\doibase 10.1103/physrevlett.88.166401} {\bibfield
  {journal} {\bibinfo  {journal} {Phys. Rev. Lett.}\ }\textbf {\bibinfo
  {volume} {88}},\ \bibinfo {pages} {166401} (\bibinfo {year}
  {2002})}\BibitemShut {NoStop}%
\bibitem [{\citenamefont {Savrasov}\ and\ \citenamefont
  {Kotliar}(2004)}]{Savrasov2004PhysRevB}%
  \BibitemOpen
  \bibfield  {author} {\bibinfo {author} {\bibfnamefont {S.~Y.}\ \bibnamefont
  {Savrasov}}\ and\ \bibinfo {author} {\bibfnamefont {G.}~\bibnamefont
  {Kotliar}},\ }\href {\doibase 10.1103/physrevb.69.245101} {\bibfield
  {journal} {\bibinfo  {journal} {Phys. Rev. B}\ }\textbf {\bibinfo {volume}
  {69}},\ \bibinfo {pages} {245101} (\bibinfo {year} {2004})}\BibitemShut
  {NoStop}%
\bibitem [{\citenamefont {Anderson}(1961)}]{Anderson1961PhysRev}%
  \BibitemOpen
  \bibfield  {author} {\bibinfo {author} {\bibfnamefont {P.~W.}\ \bibnamefont
  {Anderson}},\ }\href {\doibase 10.1103/physrev.124.41} {\bibfield  {journal}
  {\bibinfo  {journal} {Phys. Rev.}\ }\textbf {\bibinfo {volume} {124}},\
  \bibinfo {pages} {41–53} (\bibinfo {year} {1961})}\BibitemShut {NoStop}%
\bibitem [{\citenamefont {Yosida}\ and\ \citenamefont
  {Yamada}(1970)}]{Yosida1970ProgressofTheoreticalPhysicsSupplement}%
  \BibitemOpen
  \bibfield  {author} {\bibinfo {author} {\bibfnamefont {K.}~\bibnamefont
  {Yosida}}\ and\ \bibinfo {author} {\bibfnamefont {K.}~\bibnamefont
  {Yamada}},\ }\href {\doibase 10.1143/ptps.46.244} {\bibfield  {journal}
  {\bibinfo  {journal} {Progress of Theoretical Physics Supplement}\ }\textbf
  {\bibinfo {volume} {46}},\ \bibinfo {pages} {244–255} (\bibinfo {year}
  {1970})}\BibitemShut {NoStop}%
\bibitem [{\citenamefont
  {Yamada}(1975)}]{Yamada1975ProgressofTheoreticalPhysics}%
  \BibitemOpen
  \bibfield  {author} {\bibinfo {author} {\bibfnamefont {K.}~\bibnamefont
  {Yamada}},\ }\href {\doibase 10.1143/ptp.53.970} {\bibfield  {journal}
  {\bibinfo  {journal} {Progress of Theoretical Physics}\ }\textbf {\bibinfo
  {volume} {53}},\ \bibinfo {pages} {970–986} (\bibinfo {year}
  {1975})}\BibitemShut {NoStop}%
\bibitem [{\citenamefont {Horvati\'c}\ \emph {et~al.}(1987)\citenamefont
  {Horvati\'c}, \citenamefont {Sokcevi\'c},\ and\ \citenamefont
  {Zlati\'c}}]{Horvatic1987PhysRevB}%
  \BibitemOpen
  \bibfield  {author} {\bibinfo {author} {\bibfnamefont {B.}~\bibnamefont
  {Horvati\'c}}, \bibinfo {author} {\bibfnamefont {D.}~\bibnamefont
  {Sokcevi\'c}}, \ and\ \bibinfo {author} {\bibfnamefont {V.}~\bibnamefont
  {Zlati\'c}},\ }\href {\doibase 10.1103/physrevb.36.675} {\bibfield  {journal}
  {\bibinfo  {journal} {Phys. Rev. B}\ }\textbf {\bibinfo {volume} {36}},\
  \bibinfo {pages} {675-683} (\bibinfo {year} {1987})}\BibitemShut {NoStop}%
\bibitem [{\citenamefont {Kozik}\ \emph {et~al.}(2015)\citenamefont {Kozik},
  \citenamefont {Ferrero},\ and\ \citenamefont
  {Georges}}]{Kozik2015PhysRevLett}%
  \BibitemOpen
  \bibfield  {author} {\bibinfo {author} {\bibfnamefont {E.}~\bibnamefont
  {Kozik}}, \bibinfo {author} {\bibfnamefont {M.}~\bibnamefont {Ferrero}}, \
  and\ \bibinfo {author} {\bibfnamefont {A.}~\bibnamefont {Georges}},\ }\href
  {\doibase 10.1103/physrevlett.114.156402} {\bibfield  {journal} {\bibinfo
  {journal} {Phys. Rev. Lett.}\ }\textbf {\bibinfo {volume} {114}},\ \bibinfo
  {pages} {156402} (\bibinfo {year} {2015})}\BibitemShut {NoStop}%
\bibitem [{\citenamefont {Eder}(2014)}]{Eder2014arXiv}%
  \BibitemOpen
  \bibfield  {author} {\bibinfo {author} {\bibfnamefont {R.}~\bibnamefont
  {Eder}},\ }\href {http://arxiv.org/abs/1407.6599v1} {\bibfield  {journal}
  {\bibinfo  {journal} {arXiv}\ } (\bibinfo {year} {2014})},\ \Eprint
  {http://arxiv.org/abs/1407.6599v1} {arXiv:1407.6599v1 [cond-mat.str-el]}
  \BibitemShut {NoStop}%
\bibitem [{\citenamefont {Stan}\ \emph {et~al.}(2015)\citenamefont {Stan},
  \citenamefont {Romaniello}, \citenamefont {Rigamonti}, \citenamefont
  {Reining},\ and\ \citenamefont {Berger}}]{Stan2015NewJPhys}%
  \BibitemOpen
  \bibfield  {author} {\bibinfo {author} {\bibfnamefont {A.}~\bibnamefont
  {Stan}}, \bibinfo {author} {\bibfnamefont {P.}~\bibnamefont {Romaniello}},
  \bibinfo {author} {\bibfnamefont {S.}~\bibnamefont {Rigamonti}}, \bibinfo
  {author} {\bibfnamefont {L.}~\bibnamefont {Reining}}, \ and\ \bibinfo
  {author} {\bibfnamefont {J.~A.}\ \bibnamefont {Berger}},\ }\href {\doibase
  10.1088/1367-2630/17/9/093045} {\bibfield  {journal} {\bibinfo  {journal}
  {New J. Phys.}\ }\textbf {\bibinfo {volume} {17}},\ \bibinfo {pages} {093045}
  (\bibinfo {year} {2015})}\BibitemShut {NoStop}%
\bibitem [{\citenamefont {Rossi}\ and\ \citenamefont
  {Werner}(2015)}]{Rossi2015JPhysA:MathTheor}%
  \BibitemOpen
  \bibfield  {author} {\bibinfo {author} {\bibfnamefont {R.}~\bibnamefont
  {Rossi}}\ and\ \bibinfo {author} {\bibfnamefont {F.}~\bibnamefont {Werner}},\
  }\href {\doibase 10.1088/1751-8113/48/48/485202} {\bibfield  {journal}
  {\bibinfo  {journal} {J. Phys. A: Math. Theor.}\ }\textbf {\bibinfo {volume}
  {48}},\ \bibinfo {pages} {485202} (\bibinfo {year} {2015})}\BibitemShut
  {NoStop}%
\bibitem [{\citenamefont {Sch\"afer}\ \emph {et~al.}(2016)\citenamefont
  {Schäfer}, \citenamefont {Ciuchi}, \citenamefont {Wallerberger},
  \citenamefont {Thunstr\"om}, \citenamefont {Gunnarsson}, \citenamefont
  {Sangiovanni}, \citenamefont {Rohringer},\ and\ \citenamefont
  {Toschi}}]{Schafer2016PhysRevB}%
  \BibitemOpen
  \bibfield  {author} {\bibinfo {author} {\bibfnamefont {T.}~\bibnamefont
  {Sch\"afer}}, \bibinfo {author} {\bibfnamefont {S.}~\bibnamefont {Ciuchi}},
  \bibinfo {author} {\bibfnamefont {M.}~\bibnamefont {Wallerberger}}, \bibinfo
  {author} {\bibfnamefont {P.}~\bibnamefont {Thunstr\"om}}, \bibinfo {author}
  {\bibfnamefont {O.}~\bibnamefont {Gunnarsson}}, \bibinfo {author}
  {\bibfnamefont {G.}~\bibnamefont {Sangiovanni}}, \bibinfo {author}
  {\bibfnamefont {G.}~\bibnamefont {Rohringer}}, \ and\ \bibinfo {author}
  {\bibfnamefont {A.}~\bibnamefont {Toschi}},\ }\href {\doibase
  10.1103/physrevb.94.235108} {\bibfield  {journal} {\bibinfo  {journal} {Phys.
  Rev. B}\ }\textbf {\bibinfo {volume} {94}},\ \bibinfo {pages} {235108}
  (\bibinfo {year} {2016})}\BibitemShut {NoStop}%
\bibitem [{\citenamefont {Tarantino}\ \emph {et~al.}(2017)\citenamefont
  {Tarantino}, \citenamefont {Romaniello}, \citenamefont {Berger},\ and\
  \citenamefont {Reining}}]{Tarantino2017PhysRevB}%
  \BibitemOpen
  \bibfield  {author} {\bibinfo {author} {\bibfnamefont {W.}~\bibnamefont
  {Tarantino}}, \bibinfo {author} {\bibfnamefont {P.}~\bibnamefont
  {Romaniello}}, \bibinfo {author} {\bibfnamefont {J.~A.}\ \bibnamefont
  {Berger}}, \ and\ \bibinfo {author} {\bibfnamefont {L.}~\bibnamefont
  {Reining}},\ }\href {\doibase 10.1103/physrevb.96.045124} {\bibfield
  {journal} {\bibinfo  {journal} {Phys. Rev. B}\ }\textbf {\bibinfo {volume}
  {96}},\ \bibinfo {pages} {045124} (\bibinfo {year} {2017})}\BibitemShut
  {NoStop}%
\end{thebibliography}
\end{document}